\documentclass[aps,pre,twocolumn,showpacs]{revtex4-1} 
\usepackage{graphicx,amsbsy,amsmath,epsfig,natbib,float}
\usepackage{amssymb,latexsym,wasysym,pifont,mathrsfs}
\usepackage{comment,verbatim,multirow,array,sidecap,color}
\usepackage{lineno,setspace,soul,cancel,enumerate,amsmath,mathtools}
\usepackage[normalem]{ulem}
\usepackage[percent]{overpic}
\usepackage{lipsum}

\begin{document}

\title{Mechanical regularization}

\author{Himangsu Bhaumik} \email{himangsu@campus.technion.ac.il} \thanks{Present address: Yusuf Hamied Department of Chemistry, University of Cambridge, Lensfield Road, Cambridge CB2 1EW, United Kingdom}
\author{Daniel Hexner} \email{danielhe@me.technion.ac.il} \affiliation{Faculty of Mechanical Engineering, Technion, 320000 Haifa, Israel.}

\date{\today}
 
\begin{abstract}  
Training materials through periodic drive allows to endow materials and structures with complex elastic functions. As a result of the driving, the system explores the high dimensional space of structures, ultimately converging to a structure with the desired response. However, increasing the complexity of the desired response results in ultra-slow convergence and degradation. Here, we show that by constraining the search space we are able to increase robustness, extend the maximal capacity, train responses that previously did not converge, and in some cases to accelerate convergence by many orders of magnitude. We identify the geometrical constraints that prevent the formation of spurious low-frequency modes, which are responsible for failure. We argue that these constraints are analogous to regularization used in machine learning. Our results present a unified understanding of the relations between complexity, degradation, convergence, and robustness. 

\end{abstract}

\maketitle 
\subsection*{Introduction}

The multitude of variable degrees of freedom allows various systems to perform complex tasks, including computations\cite{HavaScience95}, classification\cite{kotsiantis2006machine,ciresan2011flexible}, regulation \cite{alon2007network,barabasi2004network}, and processing of high-dimensional data. Examples include neural networks (both in-vivo, and artificial) \cite{dayan2005theoretical,hopfield1982neural}, regulatory networks \cite{alon2007network,DavidsonPNAS2005}, flow networks \cite{BhattacharyyaPRL22}, and more recently mechanical structures \cite{rocks2017designing,rocks2019limits,stern2021supervised,pashine2019directed,nidhiPRM21,nidhiSM23,patil2023selflearning,LeeScienceRobotic22}. Understanding how to adjust the multitude of microscopic degrees of freedom is a central challenge in obtaining desired functionality of complex systems. 

Having a large number of parameters typically implies that these systems are over-parameterized. That is, there is a large set of solutions, each with a different set of microscopic parameters, which in principle, could have different properties. Biasing the search algorithms could provide beneficial solutions, which for example, are more robust, or perhaps more expressive \cite{bishop1995noise,neelakantan2017adding}. Indeed, machine learning algorithms employ regularization techniques, for example, to prevent over-fitting, yielding better generalization. Examples of regularization methods include constraining the set of parameters, biasing the loss function, and early stopping \cite{Goodfellow-et-al-2016}. 

In this paper, we introduce a regularization method for training elastic responses in viscoelastic structures \cite{maxwell1867iv}. We build on recent ideas for endowing precise elastic responses in mechanical systems without the aid of a computer \cite{pashine2019directed,hexner2020periodic,anisetti2023learning,stern2021supervised,patil2023selflearning,arinze2023learning}. A material is trained by applying sequences of strains that produce changes to the structure through plastic deformations. Through repetitive driving, the system may converge to the desired response. The benefit of training materials, as opposed to design and fabrication \cite{DesignExperiment}, is that it relinquishes the need to manually control a large number of microscopic degrees of freedom. 

The notion of material training introduces new considerations \cite{bhaumikPRR2022loss}. Firstly, the time scales or cycles needed to train new responses are important factors, especially in light of recent findings that convergence can be very slow. Secondly, training requires repetitive external driving, which has the effect of degrading the material \cite{bhaumikPRR2022loss,suresh1998fatigue}. Lastly, it is desirable to find solutions that are robust to small perturbations of the structure. 

In the paper, we show that constraining the angles between bonds has a profound effect on training and acts as a regularizer. While reducing the accessible set of solutions, it selects solutions with beneficial properties. The solutions that are found have an overall larger rigidity, capacity, and increased robustness. Surprisingly, in some cases, the constrained search may accelerate convergence. We present a unified understanding, based on analysis of the density of state, of the relation between degradation, convergence, complexity, and robustness.

\subsection*{Model \& Training algorithm} We study a bonded network of Maxwell viscoelastic elements, each composed of a spring and a dashpot in series\cite{maxwell1867iv}. The tension on each bond is given by,
\begin{eqnarray}
t_i &=& k_i \left( \ell_i -\ell_{i,0} \right),
\end{eqnarray}
where $k_i$ is the spring constant, $\ell_i$ is the bond's length and $\ell_{i,0}$ is the rest length. Our goal is to control the elastic response of the network by altering the geometry of the network, through changes to the rest lengths (``learning degrees of freedom''). We assume that the rest lengths evolve through plastic deformations that depend linearly on the tension on the bond; i.e., linear dashpots, 
\begin{equation}
\partial_t \ell_{i,0}=\gamma k_i (\ell_i-\ell_{i,0})
\label{eq:dl0}
\end{equation}
Throughout this paper, we take the quasistatic limit where the time scales to reach force balance are short in comparison to the time scales associated with the evolution of the dashpots. 

The bonded networks are taken to be random (as shown in Fig. \ref{fig_schematic}(a)); details can be found in the Appendix. The elastic properties of disordered networks depend on the coordination number, $Z=\frac{2N_B}{N}$, where $N_B$ is the number of bonds, and $N$ is the number of nodes. When $Z>Z_C\approx 2d$ the networks are rigid \cite{Durian1995,Ohern,liu2010jamming}. We consider two limits, small and large excess coordination number $\Delta Z \equiv Z-Z_c$ since their elastic properties are different. For small $\Delta Z$ the network is isostatic and the elastic response is anomalously long-ranged \cite{ellenbroek2006critical,lerner2014breakdown}, whereas, for large $\Delta Z$ the elastic response approximately behaves as in continuum elasticity \cite{lerner2014breakdown}. 

 As a test-bed for regularization, we consider responses whose complexity (or difficulty) can be tuned, by varying the number of ``target'' sites whose response we wish to tune and the strain amplitude $\epsilon_{Age}$. Following Ref.\cite{rocks2019limits,bhaumikPRR2022loss}, the desired response is such that an input strain on a single source site yields a prescribed strain on a $N_T$ target site. For simplicity, the input and output strain amplitude is taken to be the same, $\epsilon_{Age}$, however, the response on each target is chosen with equal probabilities to be either in-phase or out-of-phase.

The source and target site are coupled through an energy ``valley'', as illustrated in Fig. \ref{fig_schematic}(e) and (f) along the $\epsilon_T=\epsilon_S$ direction. That is, the energy as a function of strains on the source and targets is small along the desired trajectory in comparison to the transverse directions ($\epsilon_T=-\epsilon_S$). This can be understood within linear response and is discussed below. Periodic drive along the desired motion, during training, gradually reduces the energy along that path through plastic deformations that change the rest lengths. Ultimately, training converges to the desired response, provided that the response is not too difficult \cite{bhaumikPRR2022loss}. 

With increasing complexity training becomes difficult which is expressed in the slowdown of convergence \cite{bhaumikPRR2022loss}. At a critical threshold, the convergence time appears to diverge marking the limit of trainable responses. As noted, failure occurs through the proliferation of low-frequency modes, which we refer to as degradation. We have previously found that these can be traced to local geometrical features, where pairs of bonds nearly align\cite{bhaumikPRR2022loss}. We are therefore motivated to prevent the angles between bonds from becoming small. To this end, we introduce an angular repulsive force that acts when $\theta<\theta_c$, illustrated in Fig. \ref{fig_schematic} (d) and discussed in detail in the Appendix and in the supplementary information. Fig. \ref{fig_schematic}(f) illustrates the effect of the angular repulsion to increase the transverse stiffnesses.

\begin{figure}[t]
  \centerline{
        \includegraphics[width=.95\linewidth]{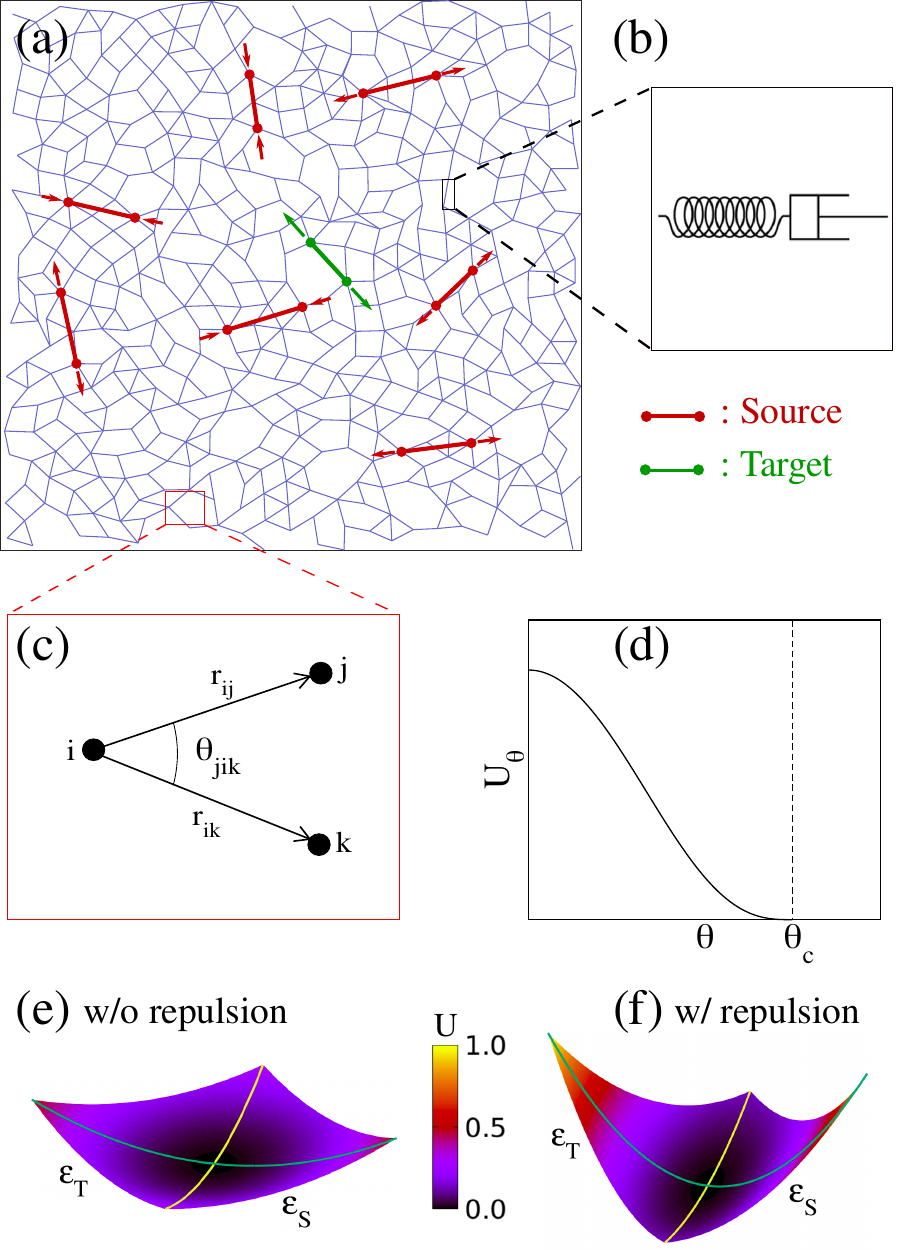}
        }
\caption{\label{fig_schematic} {\bf An illustration of the model:} (a) An example of an elastic network with $N=500$ nodes. Each pair of source (green) and target(red) sites are connected by a line, and the arrows represent the phase of the response. (b) Each bond is a spring and a dashpot in series. The dashpot allows for changes in the structure which alter the elastic properties. (c) In addition, we include an angler repulsive force that prevents the angles between adjacent bonds from becoming small. (d) An illustration of the angular potential which is non-zero only when $\theta<\theta_c$. (e) and (f) an illustration of the effect of angular regularization (repulsion). The energy valley with repulsion is deeper and has a larger transverse stiffness, which leads to enhanced properties.} 

\end{figure}

\begin{figure}[t]
  \centerline{
        \includegraphics[width=.99\linewidth]{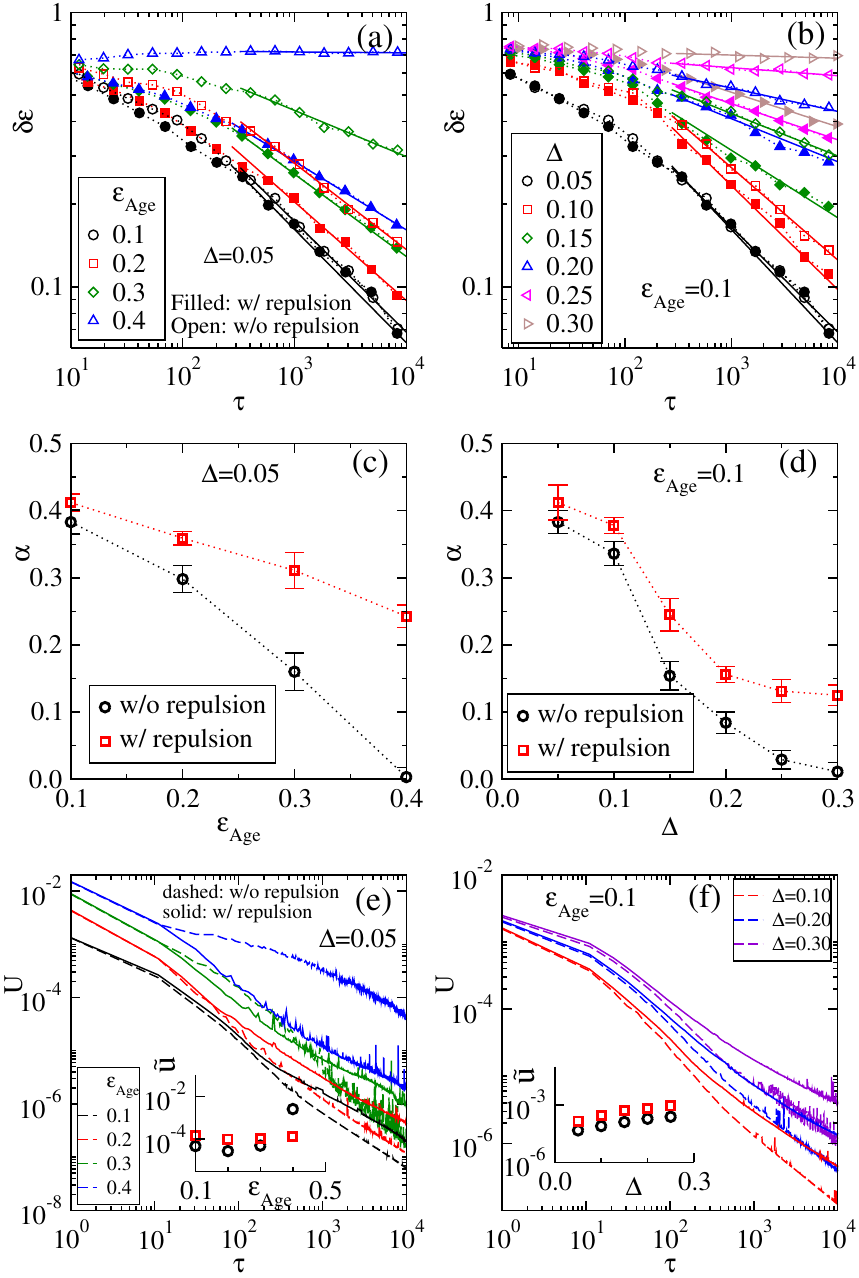}
        }
\caption{\label{fig_error} {\bf Training at small coordination number: } Left: dependence on $\epsilon_{Age}$; right: dependence on $\Delta$. (a), (b) The error against $\tau$ with repulsion (filled symbols) and without repulsion (open symbols). (c), (d) Convergence is faster (larger $\alpha)$ and with a larger capacity when angular repulsion is present. (e), (f) The elastic energy for the large part decreases at a slower rate in the presence of angular repulsion. Inset of (e) and (f) shows the variation of $\tilde{u}=U(\tau=10^4)/U(\tau=0)$ against complexity. Here, $N=200$ and $\Delta Z\approx 0.03$.
}
\end{figure}

\subsection*{Linear analysis}
Within the linear response, the energy landscape can be characterized by the eigenmodes and frequencies of the system. The relation between an applied force, $f$ and the resulting displacement $\delta x$ depends on the Hessian, $H$, which can be decomposed in the eigenmodes, $\vec{e_{\omega}}$: 
\begin{equation}
\vec{\delta  x}   =H^{-1} \vec{f}   = \sum_i{\frac{\vec{e_{\omega_i}}\cdot \vec{f} }{\omega_i^2} \vec{e_{\omega_i}} }
\end{equation}
Here, $\vec{f}$ corresponds to the force acting on the source, and its amplitude is set to provide a desired strain. 

 Since the training rule reduces the energy along the desired trajectory, the lowest frequency mode $\omega_1$ after training corresponds to the trained response. The remaining transverse modes compete with the desired response. Since the contribution of each mode is inversely proportional to its frequency squared, the response is predominantly influenced by the two lowest frequency modes, $\omega_1$ and $\omega_2$. Consequently, the magnitude of the error can be approximated by the ratio of the squared frequencies of these two modes: $\delta \epsilon \sim \frac{\omega_1^2}{\omega_2^2}$ \cite{bhaumikPRR2022loss}. To achieve successful training, it is necessary for $\omega_1$ to be small in comparison to $\omega_2$.

\begin{figure}[t]
  \centerline{
      \includegraphics[width=.99\linewidth]{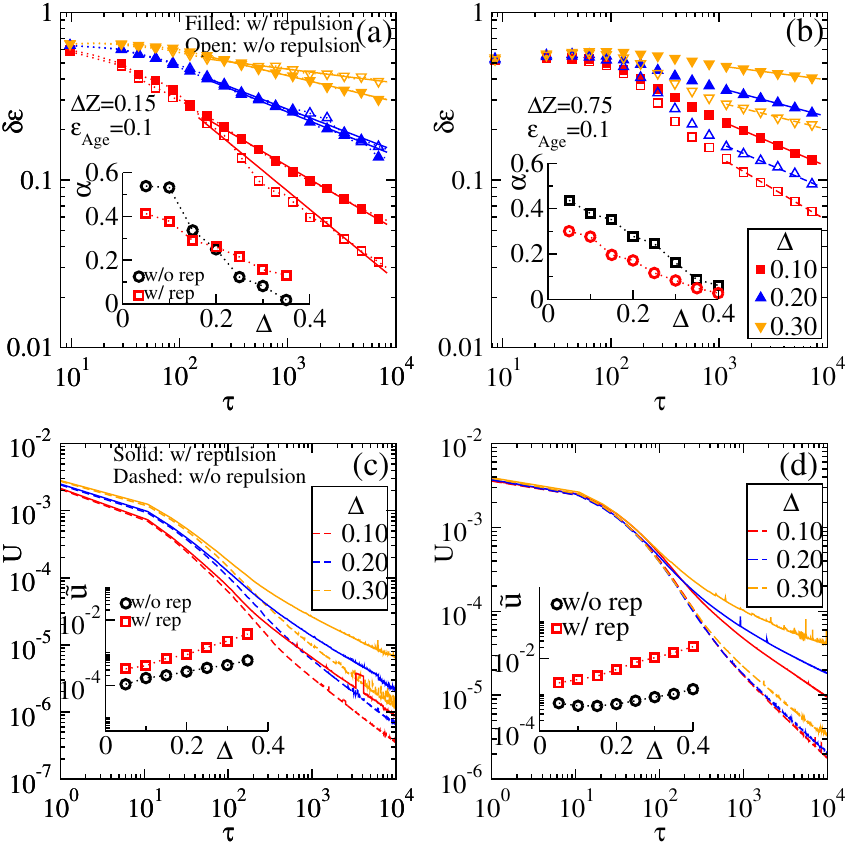}
        }
\caption{\label{fig_errorP001} {\bf Training at intermediate coordination number (left) and large coordination number (right):} (a), (b) The training error against $\tau$ for different values of $\Delta$ and at a fixed $\epsilon_{Age}$. In the inset, we show $\alpha$ which characterizes convergence. At intermediate $\Delta Z$ convergence is slower with the angular repulsion for small $\Delta$ and faster at large $\Delta$. Note, the capacity is larger with repulsion. For larger coordination numbers convergence is slower, however, the capacity is approximately the same. The energy $U$ decreases at a slower rate with repulsion. Inset of (c) and (d) shows the variation of $\tilde{u}=U(\tau=10^4)/U(\tau=0)$ against $\Delta$. Here, $N=200$. On the left $\Delta Z\approx0.15$ and on the right $\Delta Z\approx0.75$.
}
\end{figure}

\section*{Results} 
\subsection*{Accelerated convergence} 

We begin discussing the convergence of training by measuring the error between the desired and measured response on the targets, 
\begin{eqnarray}
\delta \epsilon^2=\frac{1}{N_T}\sum_T{\frac{ \left( \epsilon_{T}-\epsilon_{T}^{desired} \right)^2}{\epsilon_{Age}^2}}.
\end{eqnarray}
Here, $N_T$ is the number of targets, and $\epsilon_T$ is the strain on a target. Previously, it was found that the error approximately decays as a power-law $\tau^{-\alpha}$ and that $\alpha$ depends on the number of targets per node, denoted by, $\Delta=\frac{N_T}{N}$. With increased complexity, $\alpha$ decreases and vanishes, in what appears to be a critical point. The transition marks the maximal number of sites that can be trained \cite{bhaumikPRR2022loss}. 

We begin by considering the small $\Delta Z$ limit, shown in Fig \ref{fig_error} where we compare the results with and without the angular constraints. The left panels consider the effect of  varying the training amplitude $\epsilon_{Age}$, while in the right panels, $\Delta$ is varied. Both $\epsilon_{Age}$ and $\Delta$ are a measure of the difficulty of the trained response, and for large enough values training fails. 

The training error $\delta \epsilon $ as a function of the number of cycles, $\tau$, is shown in Figs. \ref{fig_error}(a) and (b). In both cases, convergence is approximately a power-law at large times: $\delta \epsilon\sim \tau^{-\alpha}$. The exponent $\alpha$ depends on the presence of the angular constraint as shown in Figs. \ref{fig_error}(c) and (d). The exponent decreases with increasing complexity, $\epsilon_{Age}$ or $\Delta$. Remarkably, the exponent is substantially larger for training with repulsion, implying faster convergence. This is most apparent at large $\epsilon_{Age}\approx0.4$ or $\Delta \approx0.3$, where in the absence of the angular constraints the error does not converge at all, while with the angular constrain $\alpha$ is non-zero. This implies the capacity of complex responses for training with repulsion is larger than that of training without repulsion. 

We emphasize that any small change to the exponent $\alpha$ can have an overwhelming effect on the convergence time of the training error. For example, if we train until $\delta \epsilon = \epsilon_{min}$ then the number of cycles scales as $\tau \propto \epsilon_{min}^{-1/\alpha}$. Halving, for example, the exponent reduces the convergence time to a square root of that original value. Considering the time scales for convergence near the transition is huge, and that the change in exponent exceeds a factor of two, acceleration is immense.

To track down the cause of the speedup in convergence we also measure the energy, shown in \ref{fig_error}(e) and (f). For the large part, the angular constraints reduce the rate at which energy decreases \footnote{The exception is large amplitude $\epsilon_{Age}=0.4$ in Fig.\ref{fig_error}(e). There, convergence is intermittent with sudden jumps in the energy.}. Thus, the rate of decrease in energy cannot explain the acceleration. The linear analysis suggests that acceleration is due to the suppression of transverse modes, which we further discuss below. 

Next, we study the convergence of training error as a function of $\Delta$ for larger values of the coordination number; in Fig. \ref{fig_errorP001}(a) $\Delta Z=0.15$ while in (b) $\Delta Z=0.75$. For the larger coordination number $\Delta Z=0.75$ convergence is slower, however, the capacity is weakly influenced by the angular constraints. For, $\Delta Z=0.15$ convergence is slower at small $\Delta$ and faster at larger $\Delta$. Thus, the angular constraints may accelerate or slow down convergence. 

Figs. \ref{fig_errorP001}(c) and (d) show that the angular constraints reduce the rate at which the energy decreases. We understand the slowdown or speedup as a competition between two effects. On the one hand, the energy along the desired trajectory decreases at a slower rate. On the other hand, also the stiffness along the transverse directions decreases at a slower rate. Recall, that the error scales approximately on the ratio of these two stiffnesses, as $\delta \epsilon \propto \omega_1^2/\omega_2^2$. Therefore, depending on the competition between these two effects, regularization may have the effect of either accelerating or slowing down convergence. Acceleration appears to occur at large difficulty (large $\Delta$ or large $\epsilon_{Age}$) where the excess low-frequency modes hinder convergence, as well as at small $\Delta Z$ where the density of states extends to lower frequencies, even prior to training.

\subsection*{Coupling distant sites} 
\begin{figure}[t]
  \centerline{
    \includegraphics[width=.98\linewidth]{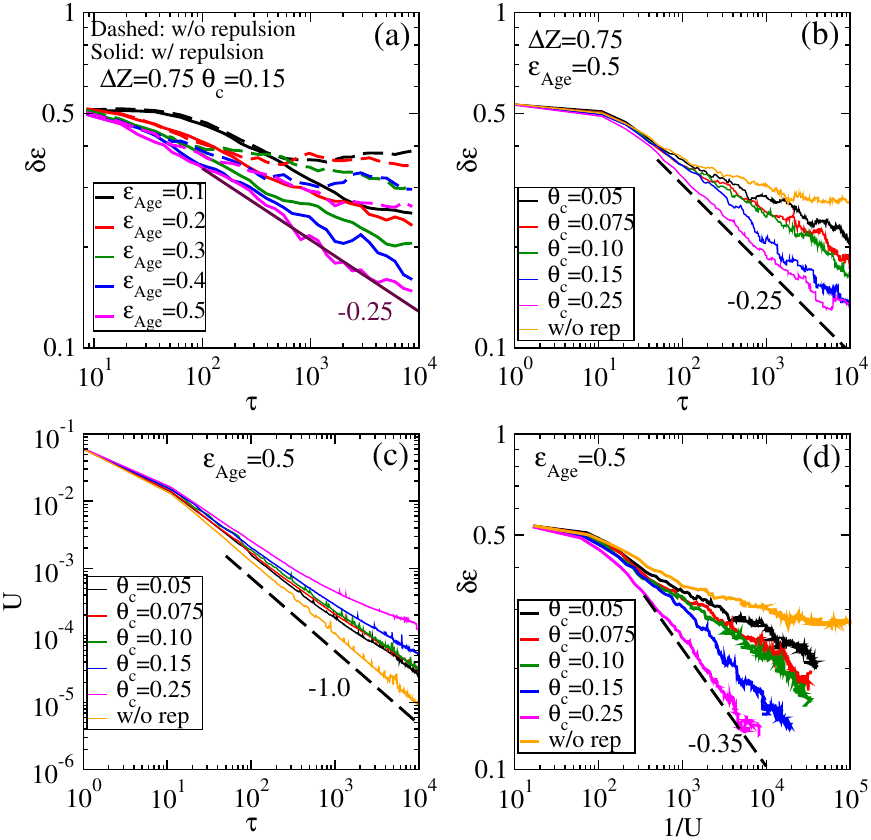} 
        }
\caption{\label{fig_allo} {\bf Training allostery inspired response:} (a) The error as a function of the number of training cycles for different amplitudes, with and without angular repulsion. Without repulsive regularization, the response does not converge. (b) The training error and (c) energy against the number of training cycles for different values of $\theta_c$. (C) The variation of the training error against the inverse of longitudinal energy $1/U$ for different $\theta_c$. Here, $N=200$, and $\Delta Z=0.75$.
}
\end{figure}
Training with the current protocol is limited by the laws of elasticity. However, here, we show that the addition of angular regularization enables responses that in their absence cannot be trained. We focus, on the allostery-inspired response, where pinching a source site results in motion on a single far-away target site \cite{rocks2017designing,yan2017architecture,eckmann2019colloquium}. 

At large $\Delta Z$, when the source and target sites are distant training does not converge. This can be understood from the elastic response to pinching a pair of nearby nodes. At large $\Delta Z$, the tension decays quickly as a function of the distance to the localized perturbation (as $r^{-d}$, where $d$ is the dimension), and as a result, distant sites do not couple \cite{hexner2020periodic}. Training fails in a manner that is different than that discussed above. Rather than creating a single low-energy mode that couples the source and target, the stiffness for actuating each individually and independently becomes small.
 
In Fig. \ref{fig_allo}(a) we show the error when training for the allostery-inspired response at different strain amplitudes, and with and without the angular repulsion. In the absence of the angular forces, the error saturates at a finite value. The error continually decreases in the presence of the angular forces, even after $10^4$ training cycles. Thus, the angular constraints enable us to train responses that do not usually converge. Interestingly, the effectiveness of the angular constraints is most pronounced at large strain amplitudes, yielding the fastest convergence. 

We also consider the effect of varying $\theta_c$. Fig. \ref{fig_allo}(b) shows that increasing $\theta_c$ results in faster convergence, leading to a smaller error. Presumably, if $\theta_c$ is too large it will hinder convergence since it limits the allowed structures. In the supplementary information, we further study the role of $\theta_c$.

Despite the faster convergence for large $\theta_c$, the energy decreases at a slower rate (see Fig.\ref{fig_allo}(c)). Presumably, increasing $\theta_c$ places a larger constraint on the allowed set of solutions and therefore slows down the rate at which energy decreases. Once again, the improvement in training is associated with preventing the transverse stiffnesses from becoming small. In Fig. \ref{fig_allo}(c) we also show the training error as a function of the inverse of longitudinal energy. The inverse energy measures the progress of training in creating an energy valley. For a given value of $U$, training is more effective with larger $\theta_c$. This measure of success also applies to training with multiple targets, and further data is presented in the supplementary information. 

\begin{figure}[t]
  \centerline{
        \includegraphics[width=.98\linewidth]
       {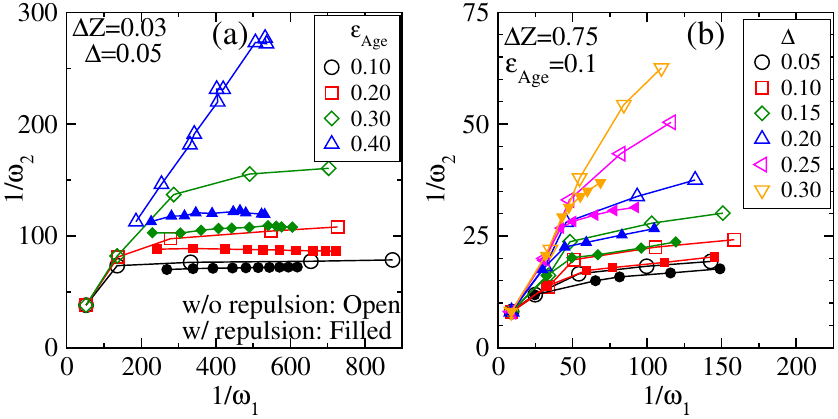}
        }
\caption{\label{fig_w1w2} {\bf Suppressed degradation:} Plot of $1/\omega_2$ against $1/\omega_1$, where $\omega_1$ is the lowest non-trivial frequency, corresponding to the trained response and $\omega_2$ is the second smallest frequency. (a) $\Delta Z \approx 0.03$ and we vary $\epsilon_{Age}$ and in (b) $\Delta Z\approx0.75$ and we vary $\Delta$. Note, that with repulsion we find solutions with a smaller $\omega_2$ per value of $\omega_1$ which leads to enhanced properties. Here, $N=200$. 
}
\end{figure}

\subsection*{Reduced degradation}
The proliferation of the spurious low-frequency modes degrades the system, reducing the overall rigidity \cite{bhaumikPRR2022loss}. Here, we show that our regularized training method arrests the spurious low-frequency modes from creeping down to low values. However, we also wish to show that it does not prevent the formation of the desired low-energy valley.

To characterize the evolution of spurious mode we measure $\omega_2$, the second lowest nontrivial eigenfrequency, per given value of lowest frequency $\omega_1$. This compares the transverse stiffness of the spurious modes to the longitudinal stiffness along the desired motion. For convenience, we plot $1/\omega_2$ as a function of $1/\omega_1$ in Fig. \ref{fig_w1w2}. Since $\omega_1$ decreases with training, $1/\omega_1$ is an indication of training time. Ideally, $1/\omega_2$ should be small in comparison to $1/\omega_1$.

Fig. \ref{fig_w1w2} shows $1/\omega_2$ vs. $1/\omega_1$ for different $\epsilon_{Age}$ and for different $\Delta$. For small $\epsilon_{Age}$ (or $\Delta$ in (b)) $1/\omega_2$ grows very slowly for both training protocols. This is the regime with low difficulty, where the training error decays faster. However, for larger values of $\epsilon_{Age}$ (or $\Delta$ in (b)), a striking difference can be observed-- $1/\omega_2$ for the training with repulsion grows much slower than that of without repulsion. 

In summary, we have shown that the angular constraints allow us to find ``solutions'' with deeper energy valleys, whose transverse stiffnesses are larger. Since there are fewer low-frequency modes, this has the effect of increasing the overall stiffness, and thus suppressing degradation. We remark that this supports our previous finding that spurious low-frequency modes originate from local structural features, where pairs of bonds align. Additional analysis regarding the spurious modes can be found in the supplementary information.

\subsection*{Enhanced robustness}
\begin{figure}[t]
  \centerline{
        \includegraphics[width=.99\linewidth]{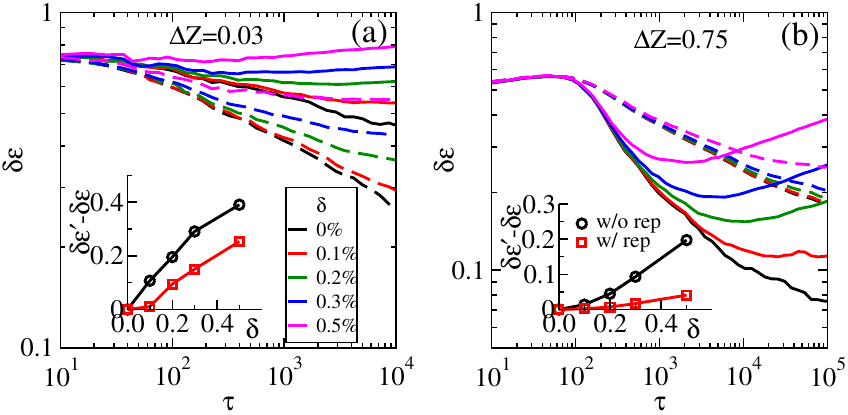}
        }
\caption{\label{fig_robustness} {\bf Robustness to small perturbations:} Training error after randomly perturbing the rest lengths with an amplitude $\delta$. In (a) $\Delta Z=0.03$ and in (b) $\Delta Z=0.75$. Dashed lines are for training with repulsion while solids are for without repulsion. Inset shows the change in error due to the perturbations, as a function of $\delta$. Note that the repulsion leads to enhanced robustness (a smaller change in the error). Here, $N=200$,$\epsilon_{Age}=0.1$, and $\Delta=0.2$.
}
\end{figure}
We next discuss the effect of the angular constraints on the robustness of the networks to small perturbations. To characterize the robustness, we alter each rest length randomly $\ell_{i,0}\rightarrow \ell_{i,0} \left (1+\delta_i \right) $, where $\delta_i$ is uniformly distributed in the range $\left[-\delta, \delta \right]$ and measure the change in the error. We emphasize that the perturbation is only a test of robustness and that the training protocol is unchanged. 

We begin by working out the predictions from the linear response. As noted, the error scales as $\omega_1^2/\omega_2^2$. The small changes to the structure change both frequencies by order $\delta$. Since, $\omega_1$ is smaller its relative change is more substantial than that of $\omega_2$, and therefore is the leading contribution to the change in the error. Note that the random changes to the structure will generically increase $\omega_1$, since it is uncharacteristically small. Therefore, the error with the perturbations will scale as, 
\begin{equation} 
\delta \epsilon^\prime \sim \frac{\omega_1^2+A \delta}{\omega_2^2} \sim \delta \epsilon+\frac{A \delta }{\omega_2^2}
\end{equation}
Here, $A$ is a constant. The change in the error is large when the transverse stiffness, $\omega_2$, is small. Therefore, a larger gap implies a higher robustness. 

In Fig. \ref{fig_robustness} we present the error as a function of the number of training cycles for different values of perturbation amplitudes, $\delta$, with and without the angular repulsion. Fig. \ref{fig_robustness}(a) shows the results for small coordination number where training is quicker with the angular repulsion. In the inset, we show that the change with repulsion is smaller, and therefore more robust.

In the case of a large coordination number, the effect is more substantial as shown in Fig. \ref{fig_robustness}(b). In the absence of any perturbation, $\delta=0$, the error is smaller when there is no angular repulsion. However, when the system is perturbed the error at long times can be smaller in the presence of angular repulsion. Thus, the angular regularization leads to more robust responses. Interestingly, for the case of no angular repulsion the error with $\delta>0$ is non-monotonic in time; initially decreases and then at later times it increases. Thus, an early stopping condition can also yield more robust results. 

\subsection*{Conclusions \& Discussion}
In summary, we have introduced a regularization method for training viscoelastic networks that constrains the angles between bonds through a repulsive force. This regularization has the effect of increasing the capacity, increasing robustness, reducing degradation, and in some cases accelerating convergence. Furthermore, it allows to train responses that previously could not be trained in the same manner. 

The central problem with the unconstrained training rule is that, while it reduces the energy along the desired trajectory, it has the unwanted effect of also reducing the energy along unintended directions \cite{bhaumikPRR2022loss}. Similar behavior was also observed when training with contrastive methods \cite{Stern2023}. Previously we traced the formation of the spurious low-frequency modes to local structural features of aligning bonds. Therefore, the angular repulsion, which constrains small angles, prevents the proliferation of soft modes, and therefore degradation. In the supplementary information, we show that this regularizer is also effective in other training rules, suggesting broader generality. 

Our analysis also provides a unified understanding of the relations between convergence, degradation, complexity, and robustness. Training is slow for complex responses because of the effect of degradation. The error depends on the ratio of the longitudinal stiffness (along the desired response) with respect to the transverse stiffness. The excess low-frequency modes compete with the desired response, requiring additional training to further reduce the energy. Degradation as we argued, also reduces robustness since the response builds on energy being very small. This makes the system particularly sensitive to small perturbations. Complex responses that have a small frequency ``gap'' are therefore less robust. 

We remark that the picture we present is remarkably similar to that of energy-based models in machine learning\cite{lecun2006tutorial}. There, information is encoded in the energy landscape, as energy minima (similarly to the Hopefield model \cite{hopfield1982neural}). Energy minima are a non-local feature of the energy landscape, necessitating its control in the high-dimensional parameter space. Contrastive methods lower the energy at the desired coordinates and raise the energy at undesirable coordinates. Regularization techniques aim at constraining the overall energy, such that lowering the energy at one set of coordinates raises the energy in the remaining landscape. This is similar to our approach, and its advantage over the contrastive approach is that it does not require directly tailoring the entire high dimensional landscape\cite{lecun2006tutorial} (or having two copies of the system \cite{dillavou2022demonstration}). 

Regularization, here, has the effect of constraining the space of solutions. In principle, the constraint may depend on the energy at other deformations. However, in that case, the training rule will depend on the behavior at different deformations, which is undesirable. Furthermore, to have relevance for training materials regularization must also be local in space, so it that can be expressed as physical local forces. It remains an open question of how to define such a local rule in other physical systems.

\section*{Appendix}
\noindent{\bf Network preparation:} The networks are constructed from a jammed configuration of repulsive soft spheres at zero temperature\cite{Ohern,Goodrich2013}. The nodes are taken to be the centers of the spheres, and overlapping spheres are attached with a bond. The advantage of this ensemble is that the coordination number is easily tuned by adjusting the applied pressure on the box. 

We select the source and target sites randomly. Each source and target site is a pair of nearby nodes, whose distance is of the order of the length of two bonds.

\noindent{\bf Angular repulsion:} Each bond of the network is assumed to be a spring with a harmonic potential $U_h=\frac{k}{2}(\ell_i-\ell_{i,0})^2$, where $\ell_i$ is the length of the bond and $\ell_{i,0}$ is the rest length. Initially, we set $\ell_i=\ell_{i,0}$, such that the network is unstressed. 

Additionally, we also consider a three-body angular repulsive potential $U_{\theta}$ that depends on the angle $\theta$ formed by the adjacent nodes. To constrain small angles, the potential is non-zero only when $\theta<\theta_c$, 
\begin{eqnarray}
U_{\theta}=\left \{ \begin{array}{l} k_{\theta}\left( 1-\frac{\cos\theta-1}{\cos\theta_c-1}\right)^3, \ \ \theta \le \theta_c \\ 0, \qquad \theta>\theta_c\end{array} \right.
\label{SIeq:Utheta}
\end{eqnarray}
where, $k_{\theta}$ is the angular stiffness, $\theta$ is the angle. In simulations, we have taken $k_{\theta}=0.1$ and verified that our results do not depend on that particular choice. We have chosen that specific potential for two reasons: 1. The angular force vanishes in a continuous manner at $\theta =\theta_c$. 2. Having a potential that depends on the cosine is easier to implement. We argue that the specific choice of potential is unimportant since its role is to constrain small angles, $\theta<\theta_c$.

\noindent{\bf Simulation methods:} We apply a local strain on source or target sites by attaching each pair of bonds with a ``ghost'' bond and varying its rest length. The strain is defined as the fractional change in their distance. Each cycle is discretized into several steps. In each step we vary the strain, minimize the energy to reach force balance, and then update the rest lengths according to equation Eq. (\ref{eq:dl0}). The energy minimization is performed using FIRE algorithm \cite{FIRE}.


\noindent{\bf Data availability:} All the relevant data for the reported work are included in the manuscript and supporting information. Additional information and data are available upon request to the corresponding author.

{\bf Acknowledgment:} {We would like to thank Marc Berneman for fruitful discussions. This work was supported by the Israel Science Foundation (grant 2385/20) and the Alon Fellowship.
}

\bibliography{ref}

\begin{thebibliography}{41}%
\makeatletter
\providecommand \@ifxundefined [1]{%
 \@ifx{#1\undefined}
}%
\providecommand \@ifnum [1]{%
 \ifnum #1\expandafter \@firstoftwo
 \else \expandafter \@secondoftwo
 \fi
}%
\providecommand \@ifx [1]{%
 \ifx #1\expandafter \@firstoftwo
 \else \expandafter \@secondoftwo
 \fi
}%
\providecommand \natexlab [1]{#1}%
\providecommand \enquote  [1]{``#1''}%
\providecommand \bibnamefont  [1]{#1}%
\providecommand \bibfnamefont [1]{#1}%
\providecommand \citenamefont [1]{#1}%
\providecommand \href@noop [0]{\@secondoftwo}%
\providecommand \href [0]{\begingroup \@sanitize@url \@href}%
\providecommand \@href[1]{\@@startlink{#1}\@@href}%
\providecommand \@@href[1]{\endgroup#1\@@endlink}%
\providecommand \@sanitize@url [0]{\catcode `\\12\catcode `\$12\catcode
  `\&12\catcode `\#12\catcode `\^12\catcode `\_12\catcode `\%12\relax}%
\providecommand \@@startlink[1]{}%
\providecommand \@@endlink[0]{}%
\providecommand \url  [0]{\begingroup\@sanitize@url \@url }%
\providecommand \@url [1]{\endgroup\@href {#1}{\urlprefix }}%
\providecommand \urlprefix  [0]{URL }%
\providecommand \Eprint [0]{\href }%
\providecommand \doibase [0]{http://dx.doi.org/}%
\providecommand \selectlanguage [0]{\@gobble}%
\providecommand \bibinfo  [0]{\@secondoftwo}%
\providecommand \bibfield  [0]{\@secondoftwo}%
\providecommand \translation [1]{[#1]}%
\providecommand \BibitemOpen [0]{}%
\providecommand \bibitemStop [0]{}%
\providecommand \bibitemNoStop [0]{.\EOS\space}%
\providecommand \EOS [0]{\spacefactor3000\relax}%
\providecommand \BibitemShut  [1]{\csname bibitem#1\endcsname}%
\let\auto@bib@innerbib\@empty
\bibitem [{\citenamefont {Siegelmann}(1995)}]{HavaScience95}%
  \BibitemOpen
  \bibfield  {author} {\bibinfo {author} {\bibfnamefont {H.~T.}\ \bibnamefont
  {Siegelmann}},\ }\href {\doibase 10.1126/science.268.5210.545} {\bibfield
  {journal} {\bibinfo  {journal} {Science}\ }\textbf {\bibinfo {volume}
  {268}},\ \bibinfo {pages} {545} (\bibinfo {year} {1995})},\ \Eprint
  {http://arxiv.org/abs/https://www.science.org/doi/pdf/10.1126/science.268.5210.545}
  {https://www.science.org/doi/pdf/10.1126/science.268.5210.545} \BibitemShut
  {NoStop}%
\bibitem [{\citenamefont {Kotsiantis}\ \emph {et~al.}(2006)\citenamefont
  {Kotsiantis}, \citenamefont {Zaharakis},\ and\ \citenamefont
  {Pintelas}}]{kotsiantis2006machine}%
  \BibitemOpen
  \bibfield  {author} {\bibinfo {author} {\bibfnamefont {S.~B.}\ \bibnamefont
  {Kotsiantis}}, \bibinfo {author} {\bibfnamefont {I.~D.}\ \bibnamefont
  {Zaharakis}}, \ and\ \bibinfo {author} {\bibfnamefont {P.~E.}\ \bibnamefont
  {Pintelas}},\ }\href@noop {} {\bibfield  {journal} {\bibinfo  {journal}
  {Artificial Intelligence Review}\ }\textbf {\bibinfo {volume} {26}},\
  \bibinfo {pages} {159} (\bibinfo {year} {2006})}\BibitemShut {NoStop}%
\bibitem [{\citenamefont {Ciresan}\ \emph {et~al.}(2011)\citenamefont
  {Ciresan}, \citenamefont {Meier}, \citenamefont {Masci}, \citenamefont
  {Gambardella},\ and\ \citenamefont {Schmidhuber}}]{ciresan2011flexible}%
  \BibitemOpen
  \bibfield  {author} {\bibinfo {author} {\bibfnamefont {D.~C.}\ \bibnamefont
  {Ciresan}}, \bibinfo {author} {\bibfnamefont {U.}~\bibnamefont {Meier}},
  \bibinfo {author} {\bibfnamefont {J.}~\bibnamefont {Masci}}, \bibinfo
  {author} {\bibfnamefont {L.~M.}\ \bibnamefont {Gambardella}}, \ and\ \bibinfo
  {author} {\bibfnamefont {J.}~\bibnamefont {Schmidhuber}},\ }in\ \href@noop {}
  {\emph {\bibinfo {booktitle} {Twenty-second international joint conference on
  artificial intelligence}}}\ (\bibinfo {organization} {Citeseer},\ \bibinfo
  {year} {2011})\BibitemShut {NoStop}%
\bibitem [{\citenamefont {Alon}(2007)}]{alon2007network}%
  \BibitemOpen
  \bibfield  {author} {\bibinfo {author} {\bibfnamefont {U.}~\bibnamefont
  {Alon}},\ }\href@noop {} {\bibfield  {journal} {\bibinfo  {journal} {Nature
  Reviews Genetics}\ }\textbf {\bibinfo {volume} {8}},\ \bibinfo {pages} {450}
  (\bibinfo {year} {2007})}\BibitemShut {NoStop}%
\bibitem [{\citenamefont {Barabasi}\ and\ \citenamefont
  {Oltvai}(2004)}]{barabasi2004network}%
  \BibitemOpen
  \bibfield  {author} {\bibinfo {author} {\bibfnamefont {A.-L.}\ \bibnamefont
  {Barabasi}}\ and\ \bibinfo {author} {\bibfnamefont {Z.~N.}\ \bibnamefont
  {Oltvai}},\ }\href@noop {} {\bibfield  {journal} {\bibinfo  {journal} {Nature
  reviews genetics}\ }\textbf {\bibinfo {volume} {5}},\ \bibinfo {pages} {101}
  (\bibinfo {year} {2004})}\BibitemShut {NoStop}%
\bibitem [{\citenamefont {Dayan}\ and\ \citenamefont
  {Abbott}(2005)}]{dayan2005theoretical}%
  \BibitemOpen
  \bibfield  {author} {\bibinfo {author} {\bibfnamefont {P.}~\bibnamefont
  {Dayan}}\ and\ \bibinfo {author} {\bibfnamefont {L.~F.}\ \bibnamefont
  {Abbott}},\ }\href@noop {} {\emph {\bibinfo {title} {Theoretical
  neuroscience: computational and mathematical modeling of neural systems}}}\
  (\bibinfo  {publisher} {MIT press},\ \bibinfo {year} {2005})\BibitemShut
  {NoStop}%
\bibitem [{\citenamefont {Hopfield}(1982)}]{hopfield1982neural}%
  \BibitemOpen
  \bibfield  {author} {\bibinfo {author} {\bibfnamefont {J.~J.}\ \bibnamefont
  {Hopfield}},\ }\href@noop {} {\bibfield  {journal} {\bibinfo  {journal}
  {Proceedings of the national academy of sciences}\ }\textbf {\bibinfo
  {volume} {79}},\ \bibinfo {pages} {2554} (\bibinfo {year}
  {1982})}\BibitemShut {NoStop}%
\bibitem [{\citenamefont {Davidson}\ and\ \citenamefont
  {Levin}(2005)}]{DavidsonPNAS2005}%
  \BibitemOpen
  \bibfield  {author} {\bibinfo {author} {\bibfnamefont {E.}~\bibnamefont
  {Davidson}}\ and\ \bibinfo {author} {\bibfnamefont {M.}~\bibnamefont
  {Levin}},\ }\href {\doibase 10.1073/pnas.0502024102} {\bibfield  {journal}
  {\bibinfo  {journal} {Proceedings of the National Academy of Sciences}\
  }\textbf {\bibinfo {volume} {102}},\ \bibinfo {pages} {4935} (\bibinfo {year}
  {2005})},\ \Eprint
  {http://arxiv.org/abs/https://www.pnas.org/doi/pdf/10.1073/pnas.0502024102}
  {https://www.pnas.org/doi/pdf/10.1073/pnas.0502024102} \BibitemShut {NoStop}%
\bibitem [{\citenamefont {Bhattacharyya}\ \emph {et~al.}(2022)\citenamefont
  {Bhattacharyya}, \citenamefont {Zwicker},\ and\ \citenamefont
  {Alim}}]{BhattacharyyaPRL22}%
  \BibitemOpen
  \bibfield  {author} {\bibinfo {author} {\bibfnamefont {K.}~\bibnamefont
  {Bhattacharyya}}, \bibinfo {author} {\bibfnamefont {D.}~\bibnamefont
  {Zwicker}}, \ and\ \bibinfo {author} {\bibfnamefont {K.}~\bibnamefont
  {Alim}},\ }\href {\doibase 10.1103/PhysRevLett.129.028101} {\bibfield
  {journal} {\bibinfo  {journal} {Phys. Rev. Lett.}\ }\textbf {\bibinfo
  {volume} {129}},\ \bibinfo {pages} {028101} (\bibinfo {year}
  {2022})}\BibitemShut {NoStop}%
\bibitem [{\citenamefont {Rocks}\ \emph {et~al.}(2017)\citenamefont {Rocks},
  \citenamefont {Pashine}, \citenamefont {Bischofberger}, \citenamefont
  {Goodrich}, \citenamefont {Liu},\ and\ \citenamefont
  {Nagel}}]{rocks2017designing}%
  \BibitemOpen
  \bibfield  {author} {\bibinfo {author} {\bibfnamefont {J.~W.}\ \bibnamefont
  {Rocks}}, \bibinfo {author} {\bibfnamefont {N.}~\bibnamefont {Pashine}},
  \bibinfo {author} {\bibfnamefont {I.}~\bibnamefont {Bischofberger}}, \bibinfo
  {author} {\bibfnamefont {C.~P.}\ \bibnamefont {Goodrich}}, \bibinfo {author}
  {\bibfnamefont {A.~J.}\ \bibnamefont {Liu}}, \ and\ \bibinfo {author}
  {\bibfnamefont {S.~R.}\ \bibnamefont {Nagel}},\ }\href@noop {} {\bibfield
  {journal} {\bibinfo  {journal} {Proceedings of the National Academy of
  Sciences}\ }\textbf {\bibinfo {volume} {114}},\ \bibinfo {pages} {2520}
  (\bibinfo {year} {2017})}\BibitemShut {NoStop}%
\bibitem [{\citenamefont {Rocks}\ \emph {et~al.}(2019)\citenamefont {Rocks},
  \citenamefont {Ronellenfitsch}, \citenamefont {Liu}, \citenamefont {Nagel},\
  and\ \citenamefont {Katifori}}]{rocks2019limits}%
  \BibitemOpen
  \bibfield  {author} {\bibinfo {author} {\bibfnamefont {J.~W.}\ \bibnamefont
  {Rocks}}, \bibinfo {author} {\bibfnamefont {H.}~\bibnamefont
  {Ronellenfitsch}}, \bibinfo {author} {\bibfnamefont {A.~J.}\ \bibnamefont
  {Liu}}, \bibinfo {author} {\bibfnamefont {S.~R.}\ \bibnamefont {Nagel}}, \
  and\ \bibinfo {author} {\bibfnamefont {E.}~\bibnamefont {Katifori}},\
  }\href@noop {} {\bibfield  {journal} {\bibinfo  {journal} {Proceedings of the
  National Academy of Sciences}\ }\textbf {\bibinfo {volume} {116}},\ \bibinfo
  {pages} {2506} (\bibinfo {year} {2019})}\BibitemShut {NoStop}%
\bibitem [{\citenamefont {Stern}\ \emph {et~al.}(2021)\citenamefont {Stern},
  \citenamefont {Hexner}, \citenamefont {Rocks},\ and\ \citenamefont
  {Liu}}]{stern2021supervised}%
  \BibitemOpen
  \bibfield  {author} {\bibinfo {author} {\bibfnamefont {M.}~\bibnamefont
  {Stern}}, \bibinfo {author} {\bibfnamefont {D.}~\bibnamefont {Hexner}},
  \bibinfo {author} {\bibfnamefont {J.~W.}\ \bibnamefont {Rocks}}, \ and\
  \bibinfo {author} {\bibfnamefont {A.~J.}\ \bibnamefont {Liu}},\ }\href@noop
  {} {\bibfield  {journal} {\bibinfo  {journal} {Physical Review X}\ }\textbf
  {\bibinfo {volume} {11}},\ \bibinfo {pages} {021045} (\bibinfo {year}
  {2021})}\BibitemShut {NoStop}%
\bibitem [{\citenamefont {Pashine}\ \emph {et~al.}(2019)\citenamefont
  {Pashine}, \citenamefont {Hexner}, \citenamefont {Liu},\ and\ \citenamefont
  {Nagel}}]{pashine2019directed}%
  \BibitemOpen
  \bibfield  {author} {\bibinfo {author} {\bibfnamefont {N.}~\bibnamefont
  {Pashine}}, \bibinfo {author} {\bibfnamefont {D.}~\bibnamefont {Hexner}},
  \bibinfo {author} {\bibfnamefont {A.~J.}\ \bibnamefont {Liu}}, \ and\
  \bibinfo {author} {\bibfnamefont {S.~R.}\ \bibnamefont {Nagel}},\ }\href@noop
  {} {\bibfield  {journal} {\bibinfo  {journal} {Science advances}\ }\textbf
  {\bibinfo {volume} {5}},\ \bibinfo {pages} {eaax4215} (\bibinfo {year}
  {2019})}\BibitemShut {NoStop}%
\bibitem [{\citenamefont {Pashine}(2021)}]{nidhiPRM21}%
  \BibitemOpen
  \bibfield  {author} {\bibinfo {author} {\bibfnamefont {N.}~\bibnamefont
  {Pashine}},\ }\href {\doibase 10.1103/PhysRevMaterials.5.065607} {\bibfield
  {journal} {\bibinfo  {journal} {Phys. Rev. Mater.}\ }\textbf {\bibinfo
  {volume} {5}},\ \bibinfo {pages} {065607} (\bibinfo {year}
  {2021})}\BibitemShut {NoStop}%
\bibitem [{\citenamefont {Pashine}\ \emph {et~al.}(2023)\citenamefont
  {Pashine}, \citenamefont {Nasab},\ and\ \citenamefont
  {Kramer-Bottiglio}}]{nidhiSM23}%
  \BibitemOpen
  \bibfield  {author} {\bibinfo {author} {\bibfnamefont {N.}~\bibnamefont
  {Pashine}}, \bibinfo {author} {\bibfnamefont {A.~M.}\ \bibnamefont {Nasab}},
  \ and\ \bibinfo {author} {\bibfnamefont {R.}~\bibnamefont
  {Kramer-Bottiglio}},\ }\href {\doibase 10.1039/D2SM01284G} {\bibfield
  {journal} {\bibinfo  {journal} {Soft Matter}\ }\textbf {\bibinfo {volume}
  {19}},\ \bibinfo {pages} {1617} (\bibinfo {year} {2023})}\BibitemShut
  {NoStop}%
\bibitem [{\citenamefont {Patil}\ \emph {et~al.}(2023)\citenamefont {Patil},
  \citenamefont {Ho},\ and\ \citenamefont {Prakash}}]{patil2023selflearning}%
  \BibitemOpen
  \bibfield  {author} {\bibinfo {author} {\bibfnamefont {V.~P.}\ \bibnamefont
  {Patil}}, \bibinfo {author} {\bibfnamefont {I.}~\bibnamefont {Ho}}, \ and\
  \bibinfo {author} {\bibfnamefont {M.}~\bibnamefont {Prakash}},\ }\href@noop
  {} {\enquote {\bibinfo {title} {Self-learning mechanical circuits},}\ }
  (\bibinfo {year} {2023}),\ \Eprint {http://arxiv.org/abs/2304.08711}
  {arXiv:2304.08711 [cond-mat.soft]} \BibitemShut {NoStop}%
\bibitem [{\citenamefont {Lee}\ \emph {et~al.}(2022)\citenamefont {Lee},
  \citenamefont {Mulder},\ and\ \citenamefont {Hopkins}}]{LeeScienceRobotic22}%
  \BibitemOpen
  \bibfield  {author} {\bibinfo {author} {\bibfnamefont {R.~H.}\ \bibnamefont
  {Lee}}, \bibinfo {author} {\bibfnamefont {E.~A.~B.}\ \bibnamefont {Mulder}},
  \ and\ \bibinfo {author} {\bibfnamefont {J.~B.}\ \bibnamefont {Hopkins}},\
  }\href {\doibase 10.1126/scirobotics.abq7278} {\bibfield  {journal} {\bibinfo
   {journal} {Science Robotics}\ }\textbf {\bibinfo {volume} {7}},\ \bibinfo
  {pages} {eabq7278} (\bibinfo {year} {2022})}\BibitemShut {NoStop}%
\bibitem [{\citenamefont {Bishop}(1995)}]{bishop1995noise}%
  \BibitemOpen
  \bibfield  {author} {\bibinfo {author} {\bibfnamefont {C.~M.}\ \bibnamefont
  {Bishop}},\ }\href@noop {} {\bibfield  {journal} {\bibinfo  {journal} {Neural
  computation}\ }\textbf {\bibinfo {volume} {7}},\ \bibinfo {pages} {108}
  (\bibinfo {year} {1995})}\BibitemShut {NoStop}%
\bibitem [{\citenamefont {Neelakantan}\ \emph {et~al.}(2017)\citenamefont
  {Neelakantan}, \citenamefont {Vilnis}, \citenamefont {Le}, \citenamefont
  {Kaiser}, \citenamefont {Kurach}, \citenamefont {Sutskever},\ and\
  \citenamefont {Martens}}]{neelakantan2017adding}%
  \BibitemOpen
  \bibfield  {author} {\bibinfo {author} {\bibfnamefont {A.}~\bibnamefont
  {Neelakantan}}, \bibinfo {author} {\bibfnamefont {L.}~\bibnamefont {Vilnis}},
  \bibinfo {author} {\bibfnamefont {Q.~V.}\ \bibnamefont {Le}}, \bibinfo
  {author} {\bibfnamefont {L.}~\bibnamefont {Kaiser}}, \bibinfo {author}
  {\bibfnamefont {K.}~\bibnamefont {Kurach}}, \bibinfo {author} {\bibfnamefont
  {I.}~\bibnamefont {Sutskever}}, \ and\ \bibinfo {author} {\bibfnamefont
  {J.}~\bibnamefont {Martens}},\ }\href
  {https://openreview.net/forum?id=rkjZ2Pcxe} {\enquote {\bibinfo {title}
  {Adding gradient noise improves learning for very deep networks},}\ }
  (\bibinfo {year} {2017})\BibitemShut {NoStop}%
\bibitem [{\citenamefont {Goodfellow}\ \emph {et~al.}(2016)\citenamefont
  {Goodfellow}, \citenamefont {Bengio},\ and\ \citenamefont
  {Courville}}]{Goodfellow-et-al-2016}%
  \BibitemOpen
  \bibfield  {author} {\bibinfo {author} {\bibfnamefont {I.}~\bibnamefont
  {Goodfellow}}, \bibinfo {author} {\bibfnamefont {Y.}~\bibnamefont {Bengio}},
  \ and\ \bibinfo {author} {\bibfnamefont {A.}~\bibnamefont {Courville}},\
  }\href@noop {} {\emph {\bibinfo {title} {Deep Learning}}}\ (\bibinfo
  {publisher} {MIT Press},\ \bibinfo {year} {2016})\ \bibinfo {note}
  {\url{http://www.deeplearningbook.org}}\BibitemShut {NoStop}%
\bibitem [{\citenamefont {Maxwell}(1867)}]{maxwell1867iv}%
  \BibitemOpen
  \bibfield  {author} {\bibinfo {author} {\bibfnamefont {J.~C.}\ \bibnamefont
  {Maxwell}},\ }\href@noop {} {\bibfield  {journal} {\bibinfo  {journal}
  {Philosophical transactions of the Royal Society of London}\ ,\ \bibinfo
  {pages} {49}} (\bibinfo {year} {1867})}\BibitemShut {NoStop}%
\bibitem [{\citenamefont {Hexner}\ \emph {et~al.}(2020)\citenamefont {Hexner},
  \citenamefont {Liu},\ and\ \citenamefont {Nagel}}]{hexner2020periodic}%
  \BibitemOpen
  \bibfield  {author} {\bibinfo {author} {\bibfnamefont {D.}~\bibnamefont
  {Hexner}}, \bibinfo {author} {\bibfnamefont {A.~J.}\ \bibnamefont {Liu}}, \
  and\ \bibinfo {author} {\bibfnamefont {S.~R.}\ \bibnamefont {Nagel}},\
  }\href@noop {} {\bibfield  {journal} {\bibinfo  {journal} {Proceedings of the
  National Academy of Sciences}\ }\textbf {\bibinfo {volume} {117}},\ \bibinfo
  {pages} {31690} (\bibinfo {year} {2020})}\BibitemShut {NoStop}%
\bibitem [{\citenamefont {Anisetti}\ \emph {et~al.}(2023)\citenamefont
  {Anisetti}, \citenamefont {Scellier},\ and\ \citenamefont
  {Schwarz}}]{anisetti2023learning}%
  \BibitemOpen
  \bibfield  {author} {\bibinfo {author} {\bibfnamefont {V.~R.}\ \bibnamefont
  {Anisetti}}, \bibinfo {author} {\bibfnamefont {B.}~\bibnamefont {Scellier}},
  \ and\ \bibinfo {author} {\bibfnamefont {J.~M.}\ \bibnamefont {Schwarz}},\
  }\href@noop {} {\bibfield  {journal} {\bibinfo  {journal} {Physical Review
  Research}\ }\textbf {\bibinfo {volume} {5}},\ \bibinfo {pages} {023024}
  (\bibinfo {year} {2023})}\BibitemShut {NoStop}%
\bibitem [{\citenamefont {Arinze}\ \emph {et~al.}(2023)\citenamefont {Arinze},
  \citenamefont {Stern}, \citenamefont {Nagel},\ and\ \citenamefont
  {Murugan}}]{arinze2023learning}%
  \BibitemOpen
  \bibfield  {author} {\bibinfo {author} {\bibfnamefont {C.}~\bibnamefont
  {Arinze}}, \bibinfo {author} {\bibfnamefont {M.}~\bibnamefont {Stern}},
  \bibinfo {author} {\bibfnamefont {S.~R.}\ \bibnamefont {Nagel}}, \ and\
  \bibinfo {author} {\bibfnamefont {A.}~\bibnamefont {Murugan}},\ }\href@noop
  {} {\bibfield  {journal} {\bibinfo  {journal} {Physical Review E}\ }\textbf
  {\bibinfo {volume} {107}},\ \bibinfo {pages} {025001} (\bibinfo {year}
  {2023})}\BibitemShut {NoStop}%
\bibitem [{\citenamefont {Thomke}(1998)}]{DesignExperiment}%
  \BibitemOpen
  \bibfield  {author} {\bibinfo {author} {\bibfnamefont {S.~H.}\ \bibnamefont
  {Thomke}},\ }\href {http://www.jstor.org/stable/2634644} {\bibfield
  {journal} {\bibinfo  {journal} {Management Science}\ }\textbf {\bibinfo
  {volume} {44}},\ \bibinfo {pages} {743} (\bibinfo {year} {1998})}\BibitemShut
  {NoStop}%
\bibitem [{\citenamefont {Bhaumik}\ and\ \citenamefont
  {Hexner}(2022)}]{bhaumikPRR2022loss}%
  \BibitemOpen
  \bibfield  {author} {\bibinfo {author} {\bibfnamefont {H.}~\bibnamefont
  {Bhaumik}}\ and\ \bibinfo {author} {\bibfnamefont {D.}~\bibnamefont
  {Hexner}},\ }\href {\doibase 10.1103/PhysRevResearch.4.L042044} {\bibfield
  {journal} {\bibinfo  {journal} {Phys. Rev. Research}\ }\textbf {\bibinfo
  {volume} {4}},\ \bibinfo {pages} {L042044} (\bibinfo {year}
  {2022})}\BibitemShut {NoStop}%
\bibitem [{\citenamefont {Suresh}(1998)}]{suresh1998fatigue}%
  \BibitemOpen
  \bibfield  {author} {\bibinfo {author} {\bibfnamefont {S.}~\bibnamefont
  {Suresh}},\ }\href@noop {} {\emph {\bibinfo {title} {Fatigue of materials}}}\
  (\bibinfo  {publisher} {Cambridge university press},\ \bibinfo {year}
  {1998})\BibitemShut {NoStop}%
\bibitem [{\citenamefont {Durian}(1995)}]{Durian1995}%
  \BibitemOpen
  \bibfield  {author} {\bibinfo {author} {\bibfnamefont {D.~J.}\ \bibnamefont
  {Durian}},\ }\href {https://link.aps.org/doi/10.1103/PhysRevLett.75.4780}
  {\bibfield  {journal} {\bibinfo  {journal} {Phys. Rev. Lett.}\ }\textbf
  {\bibinfo {volume} {75}},\ \bibinfo {pages} {4780} (\bibinfo {year}
  {1995})}\BibitemShut {NoStop}%
\bibitem [{\citenamefont {O'Hern}\ \emph {et~al.}(2003)\citenamefont {O'Hern},
  \citenamefont {Silbert}, \citenamefont {Liu},\ and\ \citenamefont
  {Nagel}}]{Ohern}%
  \BibitemOpen
  \bibfield  {author} {\bibinfo {author} {\bibfnamefont {C.~S.}\ \bibnamefont
  {O'Hern}}, \bibinfo {author} {\bibfnamefont {L.~E.}\ \bibnamefont {Silbert}},
  \bibinfo {author} {\bibfnamefont {A.~J.}\ \bibnamefont {Liu}}, \ and\
  \bibinfo {author} {\bibfnamefont {S.~R.}\ \bibnamefont {Nagel}},\ }\href@noop
  {} {\bibfield  {journal} {\bibinfo  {journal} {Phys. Rev. E}\ }\textbf
  {\bibinfo {volume} {68}},\ \bibinfo {pages} {011306} (\bibinfo {year}
  {2003})}\BibitemShut {NoStop}%
\bibitem [{\citenamefont {Liu}\ and\ \citenamefont
  {Nagel}(2010)}]{liu2010jamming}%
  \BibitemOpen
  \bibfield  {author} {\bibinfo {author} {\bibfnamefont {A.~J.}\ \bibnamefont
  {Liu}}\ and\ \bibinfo {author} {\bibfnamefont {S.~R.}\ \bibnamefont
  {Nagel}},\ }\href@noop {} {\bibfield  {journal} {\bibinfo  {journal} {Annu.
  Rev. Condens. Matter Phys.}\ }\textbf {\bibinfo {volume} {1}},\ \bibinfo
  {pages} {347} (\bibinfo {year} {2010})}\BibitemShut {NoStop}%
\bibitem [{\citenamefont {Ellenbroek}\ \emph {et~al.}(2006)\citenamefont
  {Ellenbroek}, \citenamefont {Somfai}, \citenamefont {van Hecke},\ and\
  \citenamefont {van Saarloos}}]{ellenbroek2006critical}%
  \BibitemOpen
  \bibfield  {author} {\bibinfo {author} {\bibfnamefont {W.~G.}\ \bibnamefont
  {Ellenbroek}}, \bibinfo {author} {\bibfnamefont {E.}~\bibnamefont {Somfai}},
  \bibinfo {author} {\bibfnamefont {M.}~\bibnamefont {van Hecke}}, \ and\
  \bibinfo {author} {\bibfnamefont {W.}~\bibnamefont {van Saarloos}},\
  }\href@noop {} {\bibfield  {journal} {\bibinfo  {journal} {Physical review
  letters}\ }\textbf {\bibinfo {volume} {97}},\ \bibinfo {pages} {258001}
  (\bibinfo {year} {2006})}\BibitemShut {NoStop}%
\bibitem [{\citenamefont {Lerner}\ \emph {et~al.}(2014)\citenamefont {Lerner},
  \citenamefont {DeGiuli}, \citenamefont {D{\"u}ring},\ and\ \citenamefont
  {Wyart}}]{lerner2014breakdown}%
  \BibitemOpen
  \bibfield  {author} {\bibinfo {author} {\bibfnamefont {E.}~\bibnamefont
  {Lerner}}, \bibinfo {author} {\bibfnamefont {E.}~\bibnamefont {DeGiuli}},
  \bibinfo {author} {\bibfnamefont {G.}~\bibnamefont {D{\"u}ring}}, \ and\
  \bibinfo {author} {\bibfnamefont {M.}~\bibnamefont {Wyart}},\ }\href@noop {}
  {\bibfield  {journal} {\bibinfo  {journal} {Soft Matter}\ }\textbf {\bibinfo
  {volume} {10}},\ \bibinfo {pages} {5085} (\bibinfo {year}
  {2014})}\BibitemShut {NoStop}%
\bibitem [{Note1()}]{Note1}%
  \BibitemOpen
  \bibinfo {note} {The exception is large amplitude $\epsilon _{Age}=0.4$ in
  Fig.\ref {fig_error}(e). There, convergence is intermittent with sudden jumps
  in the energy.}\BibitemShut {Stop}%
\bibitem [{\citenamefont {Yan}\ \emph {et~al.}(2017)\citenamefont {Yan},
  \citenamefont {Ravasio}, \citenamefont {Brito},\ and\ \citenamefont
  {Wyart}}]{yan2017architecture}%
  \BibitemOpen
  \bibfield  {author} {\bibinfo {author} {\bibfnamefont {L.}~\bibnamefont
  {Yan}}, \bibinfo {author} {\bibfnamefont {R.}~\bibnamefont {Ravasio}},
  \bibinfo {author} {\bibfnamefont {C.}~\bibnamefont {Brito}}, \ and\ \bibinfo
  {author} {\bibfnamefont {M.}~\bibnamefont {Wyart}},\ }\href@noop {}
  {\bibfield  {journal} {\bibinfo  {journal} {Proceedings of the National
  Academy of Sciences}\ }\textbf {\bibinfo {volume} {114}},\ \bibinfo {pages}
  {2526} (\bibinfo {year} {2017})}\BibitemShut {NoStop}%
\bibitem [{\citenamefont {Eckmann}\ \emph {et~al.}(2019)\citenamefont
  {Eckmann}, \citenamefont {Rougemont},\ and\ \citenamefont
  {Tlusty}}]{eckmann2019colloquium}%
  \BibitemOpen
  \bibfield  {author} {\bibinfo {author} {\bibfnamefont {J.-P.}\ \bibnamefont
  {Eckmann}}, \bibinfo {author} {\bibfnamefont {J.}~\bibnamefont {Rougemont}},
  \ and\ \bibinfo {author} {\bibfnamefont {T.}~\bibnamefont {Tlusty}},\
  }\href@noop {} {\bibfield  {journal} {\bibinfo  {journal} {Reviews of Modern
  Physics}\ }\textbf {\bibinfo {volume} {91}},\ \bibinfo {pages} {031001}
  (\bibinfo {year} {2019})}\BibitemShut {NoStop}%
\bibitem [{\citenamefont {Stern}\ \emph {et~al.}(2023)\citenamefont {Stern},
  \citenamefont {Liu},\ and\ \citenamefont {Balasubramanian}}]{Stern2023}%
  \BibitemOpen
  \bibfield  {author} {\bibinfo {author} {\bibfnamefont {M.}~\bibnamefont
  {Stern}}, \bibinfo {author} {\bibfnamefont {A.~J.}\ \bibnamefont {Liu}}, \
  and\ \bibinfo {author} {\bibfnamefont {V.}~\bibnamefont {Balasubramanian}},\
  }\href {\doibase 10.1101/2023.06.23.546243} {\bibfield  {journal} {\bibinfo
  {journal} {bioRxiv}\ } (\bibinfo {year} {2023}),\
  10.1101/2023.06.23.546243}\BibitemShut {NoStop}%
\bibitem [{\citenamefont {LeCun}\ \emph {et~al.}(2006)\citenamefont {LeCun},
  \citenamefont {Chopra}, \citenamefont {Hadsell}, \citenamefont {Ranzato},\
  and\ \citenamefont {Huang}}]{lecun2006tutorial}%
  \BibitemOpen
  \bibfield  {author} {\bibinfo {author} {\bibfnamefont {Y.}~\bibnamefont
  {LeCun}}, \bibinfo {author} {\bibfnamefont {S.}~\bibnamefont {Chopra}},
  \bibinfo {author} {\bibfnamefont {R.}~\bibnamefont {Hadsell}}, \bibinfo
  {author} {\bibfnamefont {M.}~\bibnamefont {Ranzato}}, \ and\ \bibinfo
  {author} {\bibfnamefont {F.}~\bibnamefont {Huang}},\ }\href@noop {}
  {\bibfield  {journal} {\bibinfo  {journal} {Predicting structured data}\
  }\textbf {\bibinfo {volume} {1}} (\bibinfo {year} {2006})}\BibitemShut
  {NoStop}%
\bibitem [{\citenamefont {Dillavou}\ \emph {et~al.}(2022)\citenamefont
  {Dillavou}, \citenamefont {Stern}, \citenamefont {Liu},\ and\ \citenamefont
  {Durian}}]{dillavou2022demonstration}%
  \BibitemOpen
  \bibfield  {author} {\bibinfo {author} {\bibfnamefont {S.}~\bibnamefont
  {Dillavou}}, \bibinfo {author} {\bibfnamefont {M.}~\bibnamefont {Stern}},
  \bibinfo {author} {\bibfnamefont {A.~J.}\ \bibnamefont {Liu}}, \ and\
  \bibinfo {author} {\bibfnamefont {D.~J.}\ \bibnamefont {Durian}},\
  }\href@noop {} {\bibfield  {journal} {\bibinfo  {journal} {Physical Review
  Applied}\ }\textbf {\bibinfo {volume} {18}},\ \bibinfo {pages} {014040}
  (\bibinfo {year} {2022})}\BibitemShut {NoStop}%
\bibitem [{\citenamefont {Goodrich}\ \emph {et~al.}(2013)\citenamefont
  {Goodrich}, \citenamefont {Ellenbroek},\ and\ \citenamefont
  {Liu}}]{Goodrich2013}%
  \BibitemOpen
  \bibfield  {author} {\bibinfo {author} {\bibfnamefont {C.~P.}\ \bibnamefont
  {Goodrich}}, \bibinfo {author} {\bibfnamefont {W.~G.}\ \bibnamefont
  {Ellenbroek}}, \ and\ \bibinfo {author} {\bibfnamefont {A.~J.}\ \bibnamefont
  {Liu}},\ }\href {\doibase 10.1039/C3SM51095F} {\bibfield  {journal} {\bibinfo
   {journal} {Soft Matter}\ }\textbf {\bibinfo {volume} {9}},\ \bibinfo {pages}
  {10993} (\bibinfo {year} {2013})}\BibitemShut {NoStop}%
\bibitem [{\citenamefont {Bitzek}\ \emph {et~al.}(2006)\citenamefont {Bitzek},
  \citenamefont {Koskinen}, \citenamefont {G\"ahler}, \citenamefont {Moseler},\
  and\ \citenamefont {Gumbsch}}]{FIRE}%
  \BibitemOpen
  \bibfield  {author} {\bibinfo {author} {\bibfnamefont {E.}~\bibnamefont
  {Bitzek}}, \bibinfo {author} {\bibfnamefont {P.}~\bibnamefont {Koskinen}},
  \bibinfo {author} {\bibfnamefont {F.}~\bibnamefont {G\"ahler}}, \bibinfo
  {author} {\bibfnamefont {M.}~\bibnamefont {Moseler}}, \ and\ \bibinfo
  {author} {\bibfnamefont {P.}~\bibnamefont {Gumbsch}},\ }\href@noop {}
  {\bibfield  {journal} {\bibinfo  {journal} {Phys. Rev. Lett.}\ }\textbf
  {\bibinfo {volume} {97}},\ \bibinfo {pages} {170201} (\bibinfo {year}
  {2006})}\BibitemShut {NoStop}%
\bibitem [{\citenamefont {Scellier}\ and\ \citenamefont
  {Bengio}(2017)}]{scellier2017equilibrium}%
  \BibitemOpen
  \bibfield  {author} {\bibinfo {author} {\bibfnamefont {B.}~\bibnamefont
  {Scellier}}\ and\ \bibinfo {author} {\bibfnamefont {Y.}~\bibnamefont
  {Bengio}},\ }\href@noop {} {\bibfield  {journal} {\bibinfo  {journal}
  {Frontiers in computational neuroscience}\ }\textbf {\bibinfo {volume}
  {11}},\ \bibinfo {pages} {24} (\bibinfo {year} {2017})}\BibitemShut {NoStop}%
\end{thebibliography}%


\begin{thebibliography}{1}
\bibitem{bhaumikPRR2022loss}
H Bhaumik, D Hexner, Loss of material trainability through an unusual
  transition.
\newblock {\em\protect\JournalTitle{Phys. Rev. Research}} \textbf{4}, L042044
  (2022).

\bibitem{stern2021supervised}
M Stern, D Hexner, JW Rocks, AJ Liu, Supervised learning in physical networks:
  From machine learning to learning machines.
\newblock {\em\protect\JournalTitle{Physical Review X}} \textbf{11}, 021045
  (2021).

\bibitem{scellier2017equilibrium}
B Scellier, Y Bengio, Equilibrium propagation: Bridging the gap between
  energy-based models and backpropagation.
\newblock {\em\protect\JournalTitle{Frontiers in computational neuroscience}}
  \textbf{11}, 24 (2017).
\end{thebibliography}
\bibliographystyle{apsrev4-1} 

\clearpage

\onecolumngrid
\section*{Supplementary Information:}
\renewcommand{\theequation}{S\arabic{equation}}
\renewcommand{\thefigure}{S\arabic{figure}}
\renewcommand{\thesection}{S-\arabic{section}} 
\renewcommand{\thesubsection}{\Alph}
\renewcommand{\thesubsection}{\Alph{subsection}}
\renewcommand{\thepage}{S\arabic{page}} 

\setcounter{subsection}{0} 
\setcounter{figure}{0} 
\setcounter{equation}{0} 
\setcounter{page}{1}

In the supplemental material, we provide the additional data and analysis that supports our findings.  The document is organized as follows: \\
{\bf Section A:} Provides details on computing the forces due to the angular potential. \\
{\bf Section B:} Shows that the angular forces act as a constraint to prevent angles from decreasing below $\theta_c$. Besides acting as a constraint the potential does not affect the behavior. \\
{\bf Section C:} Provides a normal mode analysis. \\
{\bf Section D:} Studies convergence in terms of the progression of training. \\
{\bf Section E:} Provides more details on the role of $\theta_c$. \\
{\bf Section F:} Provides evidence for the generality of the regularization rule. \\

\begin{figure}[h]
  \centerline{
    \includegraphics[width=.35\linewidth]{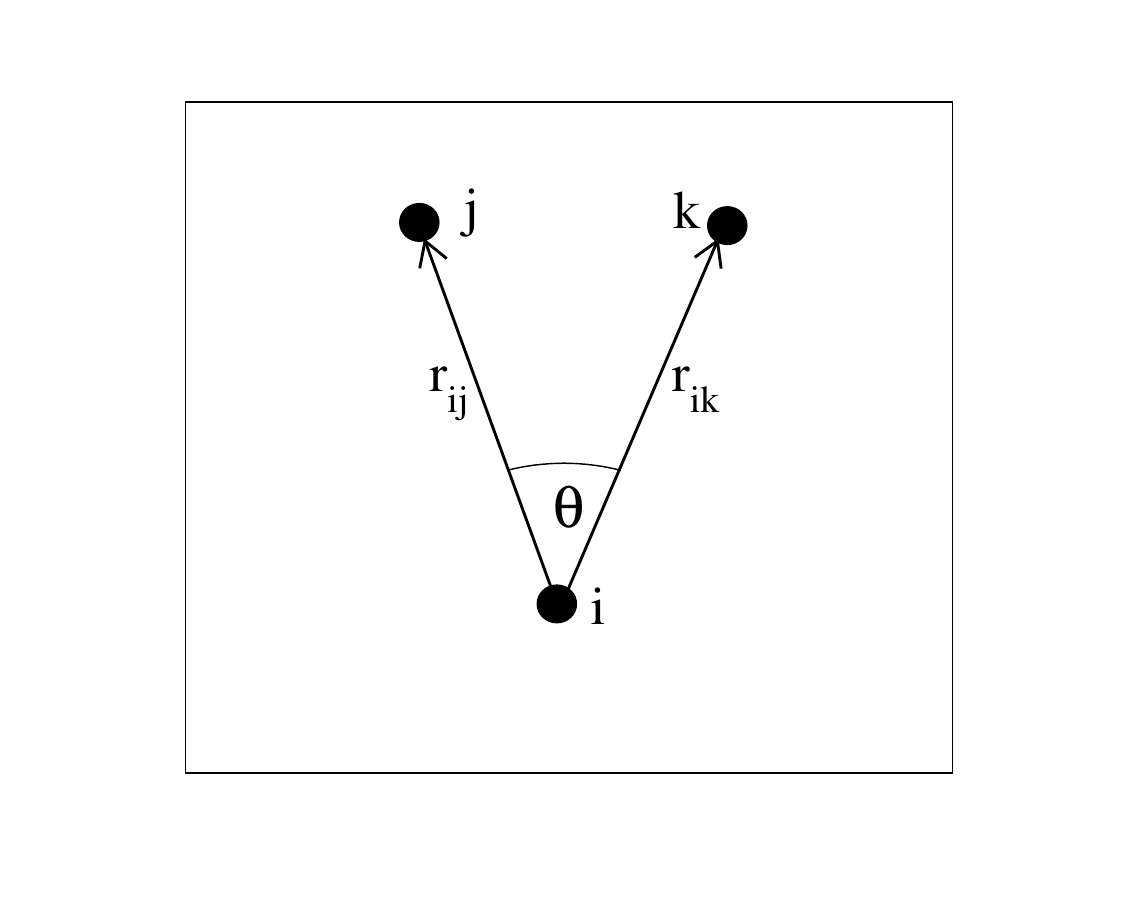}
  }
  \caption{\label{SI_fig_schem} {\bf Schematic diagram:} Representation of three nodes $i,j$, and $k$ forming an angle $\theta$ with position vector $\textbf{r}_{ij}$ and $\textbf{r}_{ik}$.}
\end{figure}

\subsection{Computation of the angular forces} Consider the schematic diagram Fig. \ref{SI_fig_schem}, where node $j$ and $k$ formed an angle $\theta$ with node $i$ by the position vector $\textbf{r}_{ij}$ and $\textbf{r}_{ik}$. In our simulations, we have taken the angular potential,
\begin{eqnarray}
  U_{\theta}=\left \{ \begin{array}{l} k_{\theta} \left( 1-\frac{\cos\theta-1}{ \cos \theta_c - 1 }\right)^3,  \ \ \theta \le \theta_c \\
    0, \qquad \theta > \theta_c \end{array} \right.
\label{SIeq:Utheta}
\end{eqnarray}

The $\beta$ Cartesian component of the force on node $l$ is derived from the angular potential, 
\begin{equation}
f_l^\beta=-\frac{\partial U_\theta}{\partial r_l^\beta}
\end{equation}
The potential,  $U_\theta$, depends on $\theta$ and therefore we employ the chain rule,
\begin{equation}
    -\frac{\partial U_\theta}{\partial r_l^\beta}= \frac{1}{\sin\theta} \frac{\partial U_\theta}{\partial \theta_{}}  \frac{\partial }{\partial r_l^\beta} \left\{  \frac{ \textbf{r}_{ij}.\textbf{r}_{ik}}{|\textbf{r}_{ij}|| \textbf{r}_{ik}| }\right\}
\end{equation}
where,
\begin{equation}
  \frac{\partial U_\theta}{\partial \theta_{}} =   k_{\theta}\left( 1-\frac{\cos\theta_{}-1}{\cos\theta_c-1}\right)^2 \frac{\sin\theta_{} }{\cos\theta_c -1}
\end{equation}
and
\begin{eqnarray}
    \frac{\partial}{\partial r_l^\beta}\left \{\frac{ \textbf{r}_{ij}.\textbf{r}_{ik}}{|\textbf{r}_{ij} ||\textbf{r}_{ik}| }\right \}&=& (\delta_{lj}-\delta_{li})\frac{r_{ik}^\beta}{|\textbf{r}_{ij}||\textbf{r}_{ik}|} + (\delta_{lk}-\delta_{li})\frac{r_{ij}^\beta}{|\textbf{r}_{ij}||\textbf{r}_{ik}|}- \nonumber\\
    &&\cos\theta_{jik} \left\{(\delta_{lj}-\delta_{li})\frac{r_{ij}^\beta}{|\textbf{r}_{ij}|^2} + (\delta_{lk}-\delta_{li})\frac{r_{ik}^\beta}{|\textbf{r}_{ik}|^2} \right\}\nonumber \\
\end{eqnarray}
Here, $\delta_{\alpha \beta}=1$ if $\alpha=\beta$ and $\delta_{\alpha \beta}=0$ otherwise. Due to Newton's third law,  the sum of the forces acting on the three nodes is zero, i.e. $f_i^\beta+f_j^\beta+f_k^\beta=0$ for each of the $\beta \in \{x,y,z\}$ components.

\subsection{\bf Angular repulsion as a constraint}
In the main text, we showed the evolution of the potential energy with the number of training cycles, $\tau$. Here we show that the angular repulsion acts as a constraint to prevent bond angles from becoming small, but otherwise has little effect on the response. That is, the energy contribution from the angular part is negligible.

\begin{figure}[h]
  \centerline{
        \includegraphics[width=.7\linewidth]{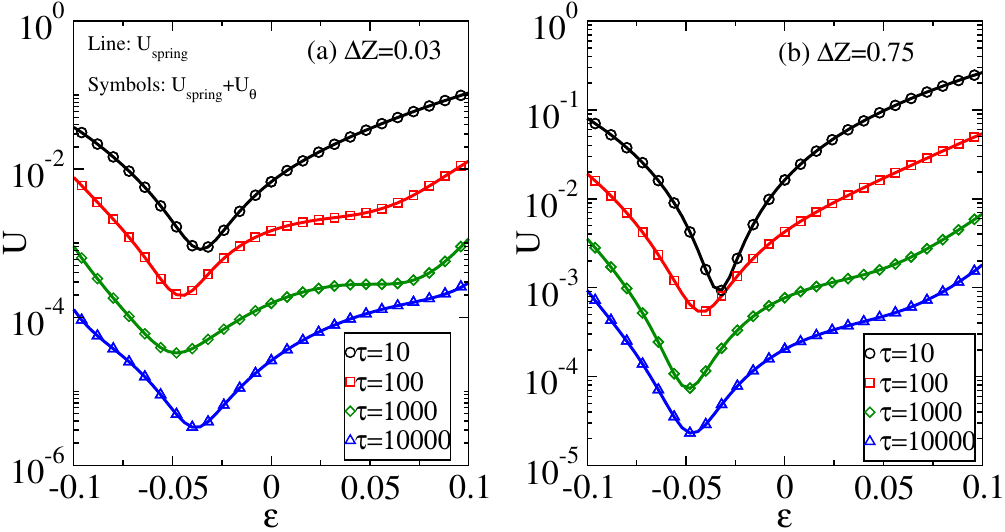}}
\caption{\label{U_overcyc_SpringAngular} The energy as a function of the strain for different numbers of training cycles ($\tau=10,10^2,10^3,$ and $10^4$) for (a) $\Delta Z=0.03$ and (b) $\Delta Z=0.75$. Lines represent the energy contributed by harmonic term only while symbols represent the energy due to harmonic spring and angular repulsion term $U_\theta$. ($\Delta=0.15$ $\epsilon_{Age}=0.1$ $N=200$)
}
\end{figure}

Figure \ref{U_overcyc_SpringAngular} illustrates the energy as a function of the strain at different stages of training for small (a)  and large (b) coordination numbers. The energy due to the harmonic (spring) potential and the angular potential are considered separately.  Remarkably, we observe that the total energy ($U_{\text{spring}}+U_\theta$), represented by the sum of the spring potential energy ($U_{\text{spring}}$) and the angular potential energy ($U_{\theta}$), aligns perfectly with the energy from the spring potential alone ($U_{\text{spring}}$). This implies that the angular contribution is negligible.

\begin{figure}[h]
  \centerline{
        \includegraphics[width=.65\linewidth]{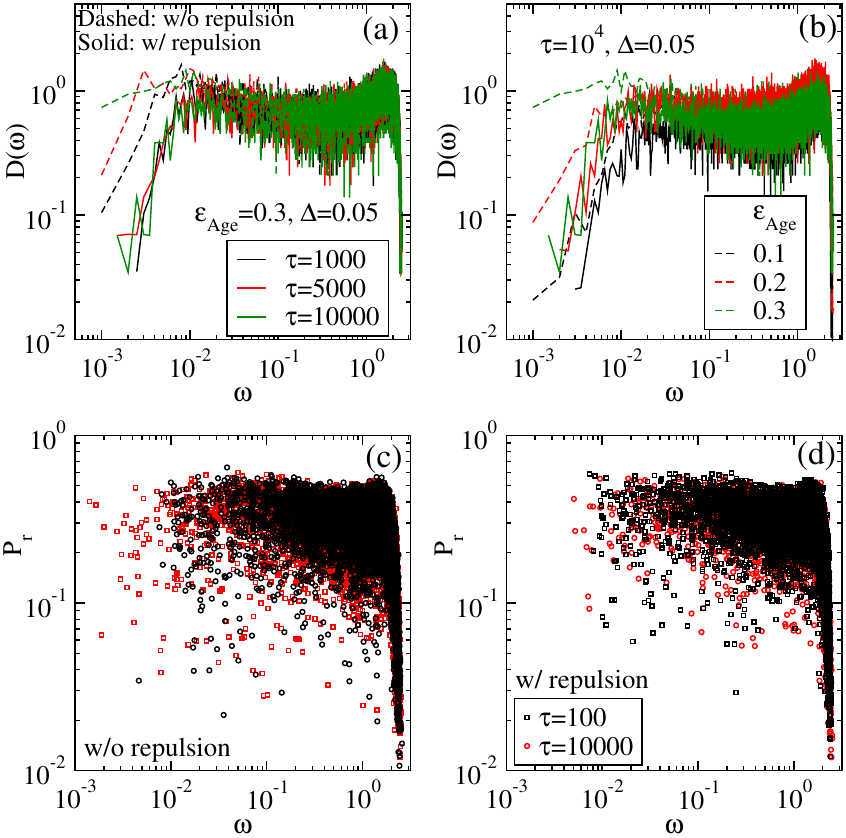}
        }
\caption{\label{SI_fig_Dw} {\bf Normal mode analysis for small coordination number, $\Delta Z=0.03$:} (a) The evolution of the density of states with the number of training cycles (dashed lines are for training without repulsion and solid lines are for training with repulsion). (b) The density of states for different strain amplitudes after $10^4$ training cycles. Here, $\Delta=0.05$. The participation ratio in the early stage($\tau=100$) and later stage ($\tau=10000$) of training. In (c) repulsion is not present and in (d) repulsion is present. Here, $N=200, \Delta Z=0.03$. 
}
\end{figure}

\subsection{\bf Effect of angular repulsion on the normal modes}
As we have shown in the current paper and previously \cite{bhaumikPRR2022loss}, the low-frequency spectrum plays an important role in the response of the system. Therefore, we study low-frequency excitations. To this end, we compute the Hessian, $H$, which is the matrix of second derivatives of the energy, and diagonalize it to find the eigenfrequencies and the eigenmodes. We characterize the low-frequency spectrum through the density of states $D(\omega)$ defined as the number of modes in the frequency range $[\omega, \omega+\delta \omega]$ per $\delta\omega$ and the number of particles. In the following, we will discuss the normal mode analysis for a small coordinated network followed by a similar analysis for a relatively highly coordinated network.

In Fig. \ref{SI_fig_Dw} we present the results of normal mode analysis for a small coordination number, $\Delta Z=0.03$.
In Fig. \ref{SI_fig_Dw} (a) we show the density of state at different stages of training with and without the angular repulsion. In Fig. \ref{SI_fig_Dw} (b) we show $D(\omega)$ after $\tau=10^4$ training cycles for different training amplitudes. In both panels, the low-frequency modes are suppressed when repulsion is present. Without repulsion, the density of states may creep down to arbitrary small values\cite{bhaumikPRR2022loss}.

\begin{figure}[t]
  \centerline{
  \includegraphics[width=.65\linewidth]{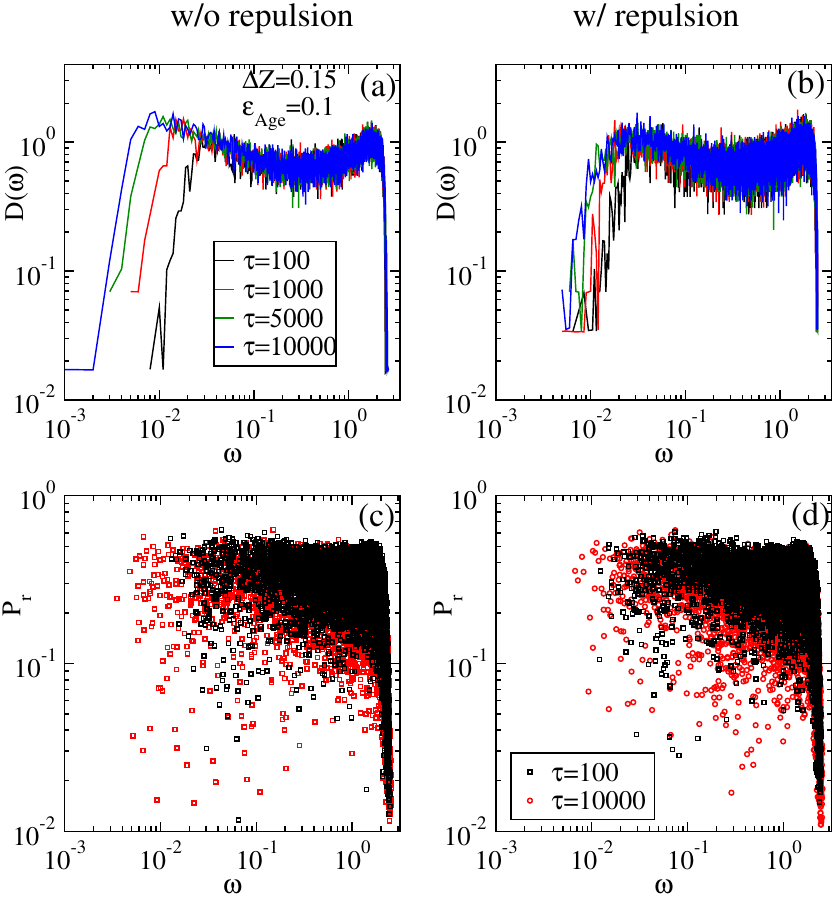}
        }     
\caption{\label{fig_SI_P001} {\bf Normal modes analysis for large coordination number,$\Delta Z\approx 0.15$:} The density of states for different numbers of training cycle, (a) training without repulsion and (b) training with repulsion. The participation ratio at different stages of training (c) without repulsion and (d) with repulsion. Here, $\Delta=0.25$.
}
\end{figure}

Previously we have found that the low-frequency modes are localized \cite{bhaumikPRR2022loss} and therefore we measure the participation ratio,

\begin{equation}
P_r=\frac{ \left( \sum_{i=1}^{N} \left| e_{r,i} \right|^2 \right)^2}{N\sum_{i=1}^{N}\left|e_{r,i}\right|^4},
\end{equation}
where $e_{r,i}$ is the normalized polarization vector of particle $i$ in the $r^{th}$ mode with frequency $\omega_r$. For an extended mode where all the particles participate equally, $P_r=1$ and for a localized mode $P_r \propto 1/N$.

In Fig. \ref{SI_fig_Dw}(c) and (d) we show the participation ratio with and without the angular repulsion respectively. We consider the behavior for small ($\tau=100$)  and large number of training cycles ($\tau=10000$). Training without repulsion results in modes with a smaller participation ratio. The spurious low-frequency modes, responsible for the failure, are localized modes with small $P_r$ values. The angular repulsion suppresses the formation of modes with a small participation ratio.

Next, we show that the effect of angular repulsion for large $\Delta Z$ is similar to that presented for low $\Delta Z$, above. 
In the absence of angular repulsion, the density of states may creep down to arbitrarily low frequencies, as suggested by Fig. \ref{fig_SI_P001}(a).  Fig. \ref{fig_SI_P001}(b) shows that angular repulsion arrests the shift to lower frequencies. In Fig. \ref{fig_SI_P001}(c) and (d) we show the participation ratio $P_r$ with and without angular repulsion. We see that similarly to the small $\Delta Z$ networks angular repulsion suppresses localized modes. 

\clearpage
\begin{figure}[!htb]
  \centerline{
    \includegraphics[width=.65\linewidth]{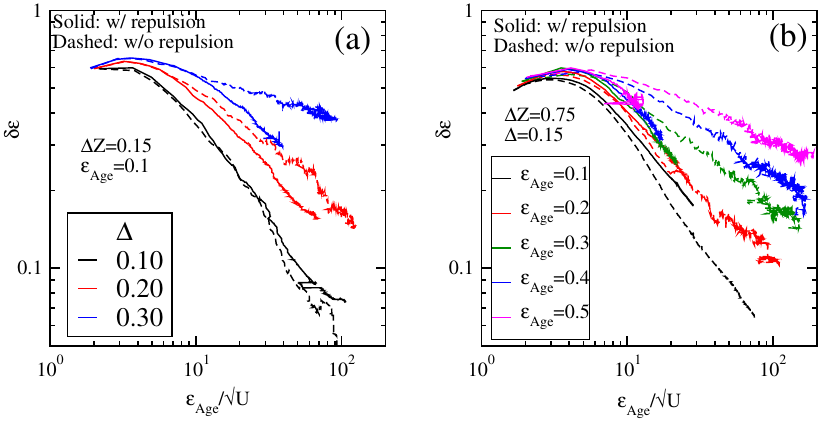}
  }
  \caption{\label{fig_SI_deps1byU_P001P01} {\bf Progression of training in terms of  energy}  (a) The training error  as a function of the inverse of the square root of energy for different numbers of target sites. The inverse energy indicates the smallness of the energy valley. Here,  $\Delta Z=0.15$ and $\epsilon_{Age}=0.1$. (b) The training error as a function of the inverse square root of the energy for different strain amplitudes. Here,  $\Delta Z=0.75$ and $\Delta=0.1$. For a better comparison, the x-axis is rescaled by the strain amplitude.}
\end{figure}
\subsection{\bf{Acceleration in terms of progression of training}}

As discussed in the main text, there are two competing factors that dictate the rate of convergence. The error depends on the ratio of the longitudinal and transverse stiffness, which in linear response scales as the ratio of the two lowest frequencies $\delta \epsilon \propto {\omega_1^2}/{\omega_2^2}$. While training creates an energy valley that coupled the source and target, it reduces the stiffness along the transverse modes. Angular constraints may accelerate or slow down convergence, depending on their effect on the two stiffnesses. 

In this section, we show that even though convergence is slower in terms of the number of cycles, it is accelerated if the progress of training is measured in terms of the smallness of the energy along the trained path. 


In Fig. \ref{fig_SI_deps1byU_P001P01} we show how training error converges with the inverse of energy for different levels of complexity (a) and strain amplitude (b). Interestingly, for a higher value of $\Delta$ or $\epsilon_{Age}$, we see a faster convergence of $\delta \epsilon$ for the training with repulsion compared to the training without repulsion. In other words, angular regularization helps the system to converge faster per inverse energy. Such acceleration is consistent with the suppression of spurious modes as presented in the main text. This also implies that the slowdown in training for large coordination numbers is due to the slower formation of the energy valley.

\clearpage
\subsection{\bf Effect of angular repulsion on the structure}

\begin{figure}[!htb]
  \centerline{
        \includegraphics[width=.6\linewidth]{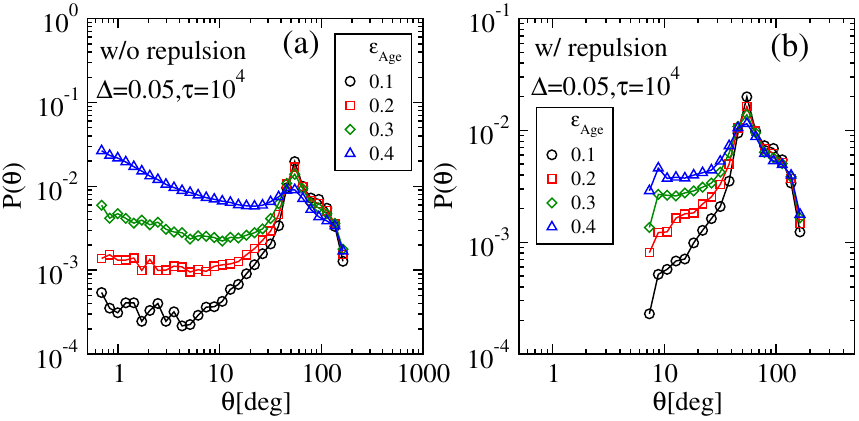}
        }
        \centerline{\quad \qquad \quad
        \includegraphics[width=.25\linewidth]{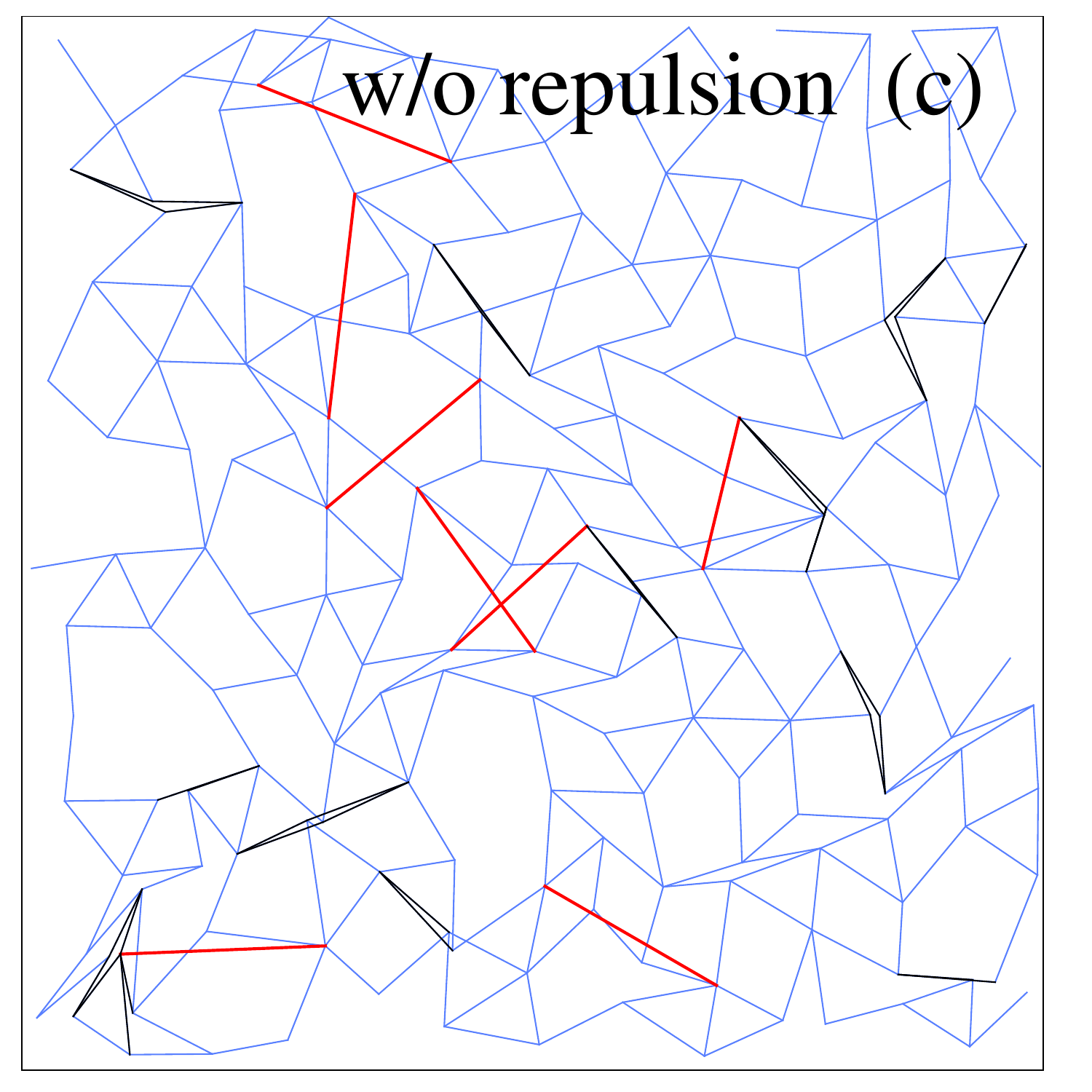}
        \qquad
        \includegraphics[width=.25\linewidth]{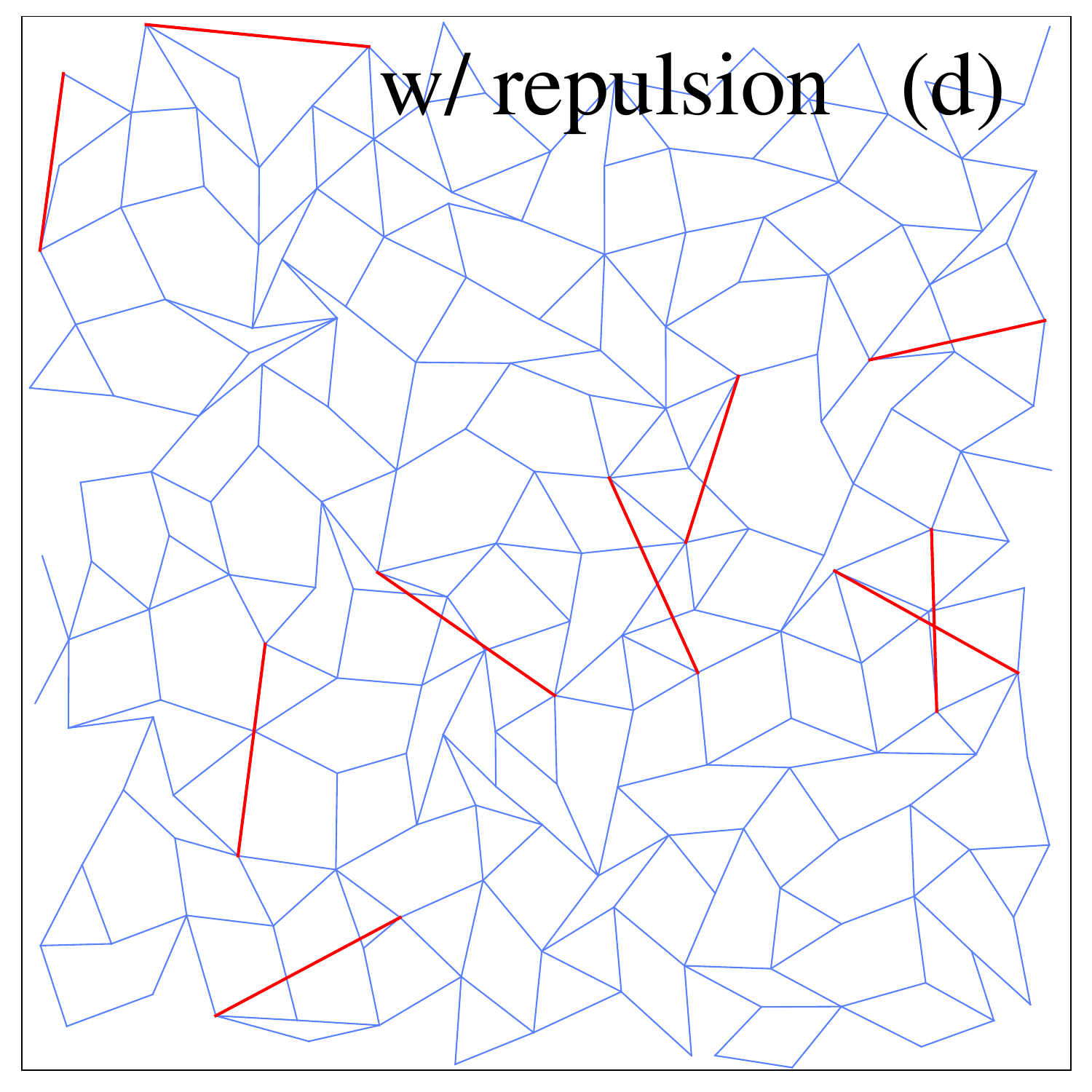}
        }
\caption{\label{SI_fig_struc} {\bf Characterizing the effect of the angular repulsion on the structure:}  Distribution of the angles between adjacent bonds for different training amplitudes, without (a) and with repulsion (b). An example of a network after it has been trained for $\tau=10^4$ cycles without (c) and with angular repulsion (d). Pairs of bonds with small angles,  $\theta< 0.15rad \ \ (\approx 8.6^o)$ are shown in black. Here, $N=200, \Delta Z=0.03$.
}
\end{figure}

Next, we study the effect of angular repulsion on the structure. Previously, it was found that the origin of the excess low-frequency modes is bonds that nearly align. That is, the angle between two bonds approaches either zero or $\pi$. In Figs. \ref{SI_fig_struc}(a) and (b) we show the distribution of angles between adjacent bonds with and without the angular repulsion, and for different values of strain amplitude. Without repulsion, the distribution at a small angle increases significantly with increasing amplitude. In contrast, with repulsion, the distribution does not decrease below the lower bound $\theta=\theta_c$. We also show the snapshot of the trained network in Fig. \ref{SI_fig_struc}(c) and (d), respectively for training without repulsion and training with repulsion. While there are many small angles formed by adjacent bonds for the training without repulsion (indicated in black), such motifs are absent for the training with repulsion. We have checked the results are similar across networks with different coordination numbers.

\clearpage

\begin{figure}[t]
  \centerline{
    \includegraphics[width=.56\linewidth]{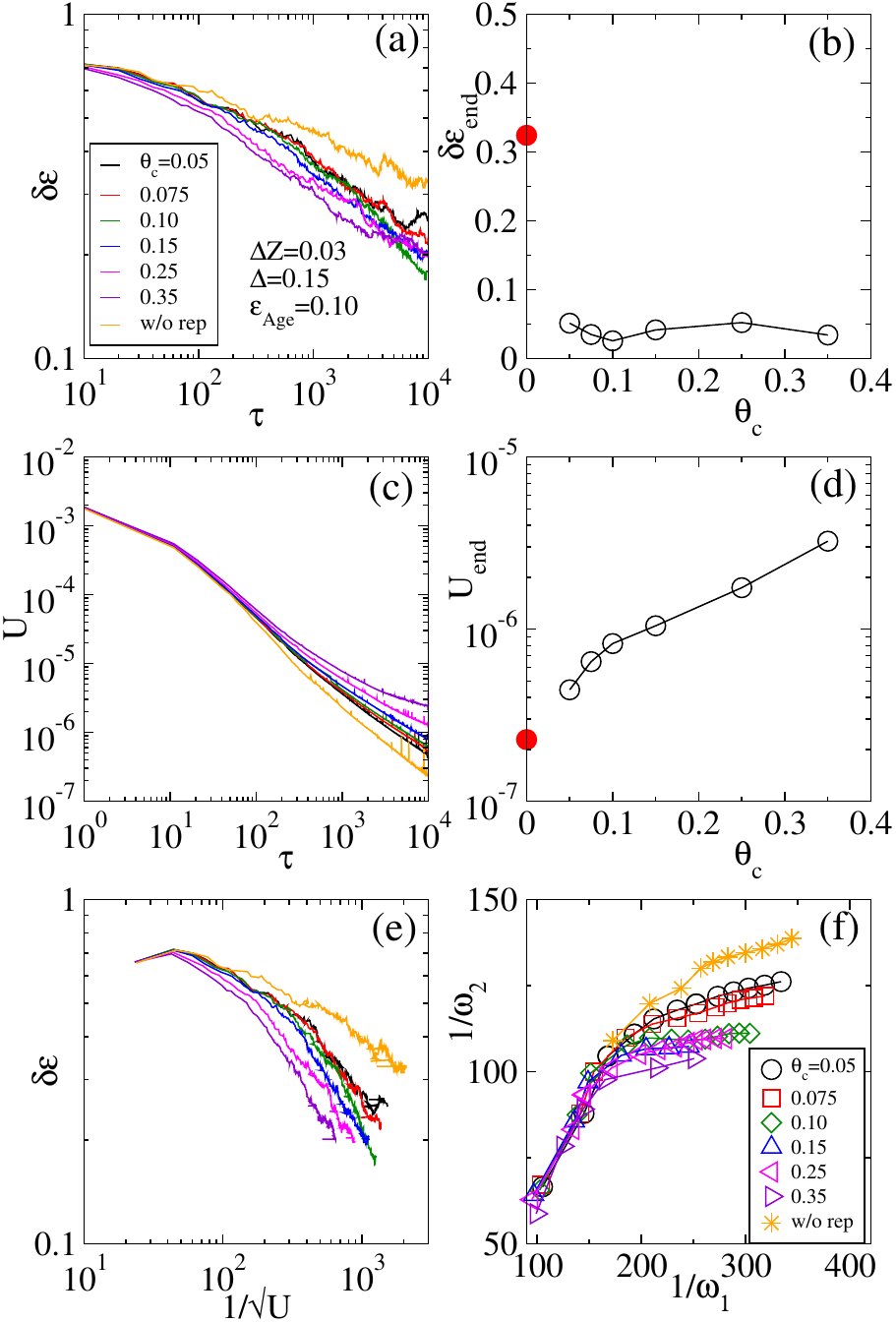}
  }
    \caption{ \label{fig_SI_thetavar_P0001} {\bf Effect of varying $\theta_c$:}  (a) The error as a function of the number of training cycles weakly depends on the value of $\theta_c$, provided it is non-zero.  (b) The error at the end of the training period ($\tau=10^4$) shows a similar effect. (c) Increasing $\theta_c$ reduces the rate at which the energy decreases. (d) The energy at the end of the training period ($\tau=10^4$) against $\theta_c$. In (b) and (d) the red dot corresponds to the absence of angular repulsion. (e) The training error against the square root of inverse energy (akin to $1/\omega_1$) converges faster with increasing $\theta_c$. (f) The spurious modes, characterized by $\omega_2$, are suppressed with increasing $\theta_c$. Here, $\Delta Z=0.03$.}
\end{figure}

\subsection{\bf Dependence on $\theta_c$} Next we study the role of critical angle $\theta_c$. We show that our results are weakly dependent on the precise value of $\theta_c$. Overall, the effect of increasing $\theta_c$ is to further restrict the formation of low-frequency modes, while also slowing down the formation of the desired energy valley. 

 Fig. \ref{fig_SI_thetavar_P0001}(a) shows the training error as a function of the number of training cycles. Fig.  \ref{fig_SI_thetavar_P0001}(b) shows the error at the end of the training period. On the scale of the simulation time, there is only a weak dependence on $\theta_c$. It appears that convergence is fastest at an intermediate value of $\theta_c$ which balances the competing effect of suppressing the spurious modes, as well as a slowdown in the formation of the energy valley. Note, that even a small $\theta_c$ fairs much better than the case of no angular constraints. 
 
In Figs. \ref{fig_SI_thetavar_P0001}(c) and (d) we show the energy as a function of the number of cycles and at the end of training against $\theta_c$. As expected, the energy decreases at a slower rate with increasing $\theta_c$. 

Fig. \ref{fig_SI_thetavar_P0001}(e) shows that convergence is accelerated when the progression of training is measured by the smallness of the energy. Increasing $\theta_c$ results in a smaller error per given value of energy along the training path. 

Lastly, we characterize in Fig. \ref{fig_SI_thetavar_P0001}(f) the suppression of the spurious low-frequency modes, as characterized by the second smallest frequency. With increasing $\theta_c$, per the given value of $\omega_1$, which corresponds to the trained response, the competing modes have a larger frequency (or stiffness). Though results presented here for $\Delta Z=0.03$, we have also verified that the qualitative behavior does not depend on the particular value of $\Delta Z$.


\subsection{Angular regularization in another training rule}
To check the generality of our regularization method we also consider another training rule, a variation of ``coupled learning'' \cite{stern2021supervised} or equilibrium propagation \cite{scellier2017equilibrium}. The evolution in coupled learning depends on a ``free state'', where a strain $\epsilon_S$ is applied to the source, and a ``clumped state'' where both the source and targets are strained. The strain applied to the targets in the clumped state is $\epsilon_T^C=\epsilon_T^F+\eta (\epsilon_D-\epsilon_T^F)$, where $\epsilon_T^F$ is the strain on the target in the free state, $\epsilon_D$ is the desired strain and  $\eta$ is the nudge factor (generally chosen to be small). As in our training rule, the responses are assumed to be quasistatic.  The learning rule is then defined as 
\begin{equation}
    \partial_t \ell_{i,0}=-\alpha \eta^{-1} k_i [\ell_i^F-\ell_i^C]
\end{equation}
where the superscript F and C denote the free and clumped state. We choose $\eta=0.1$ and the learning rate $\alpha=0.1$. It should be noted that coupled training rule may yield internal stresses. To make the rule comparable with our training rule (i.e., the resulting network is unstressed) we relax the system at the end of each training cycle, by allowing the rest length to evolve in proportion to the tension. We note that as in the main text, this rule is applied while the system is periodically strained over the desired strain range. 

\begin{figure*}[!t]
    \centerline{
    \includegraphics[width=.3\linewidth]{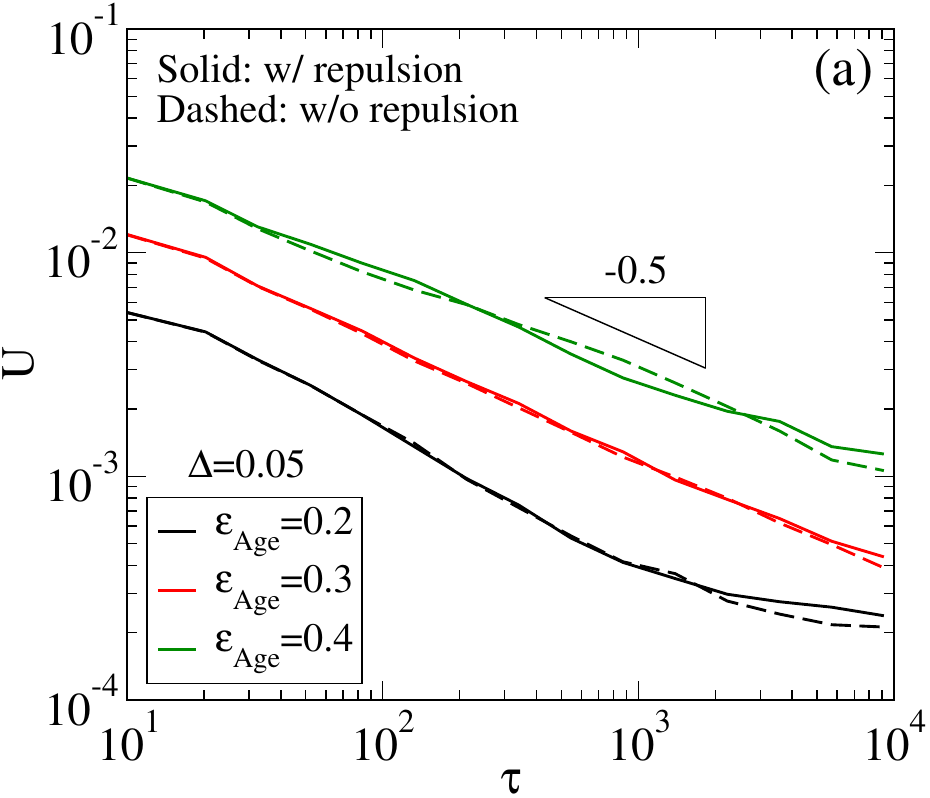}
    \includegraphics[width=.3\linewidth]{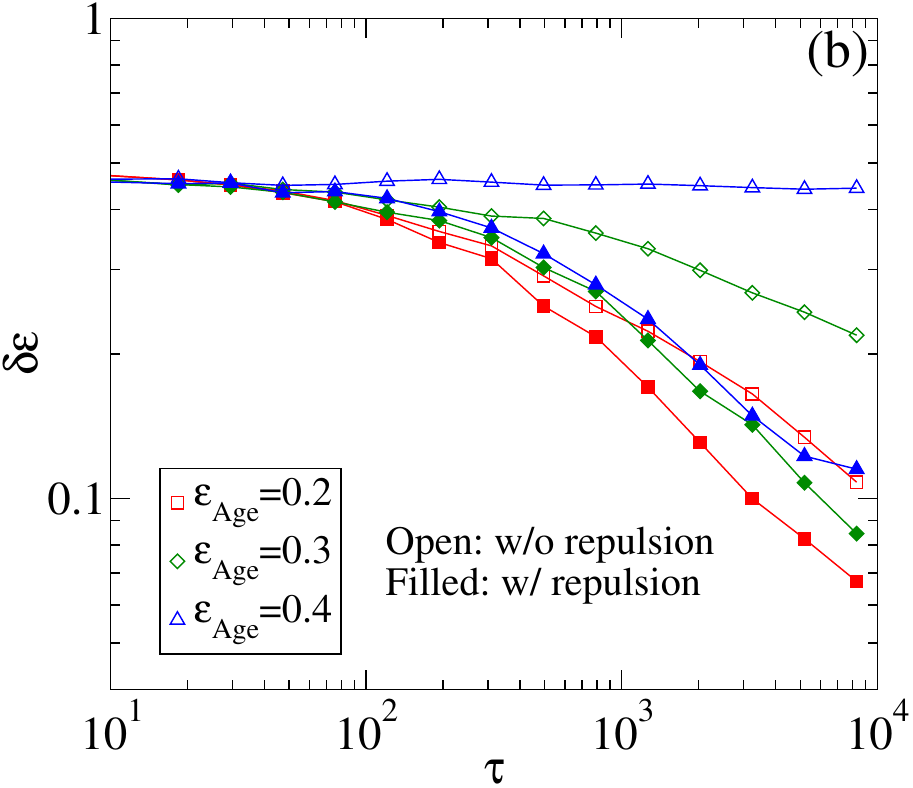}
    \includegraphics[width=.3\linewidth]{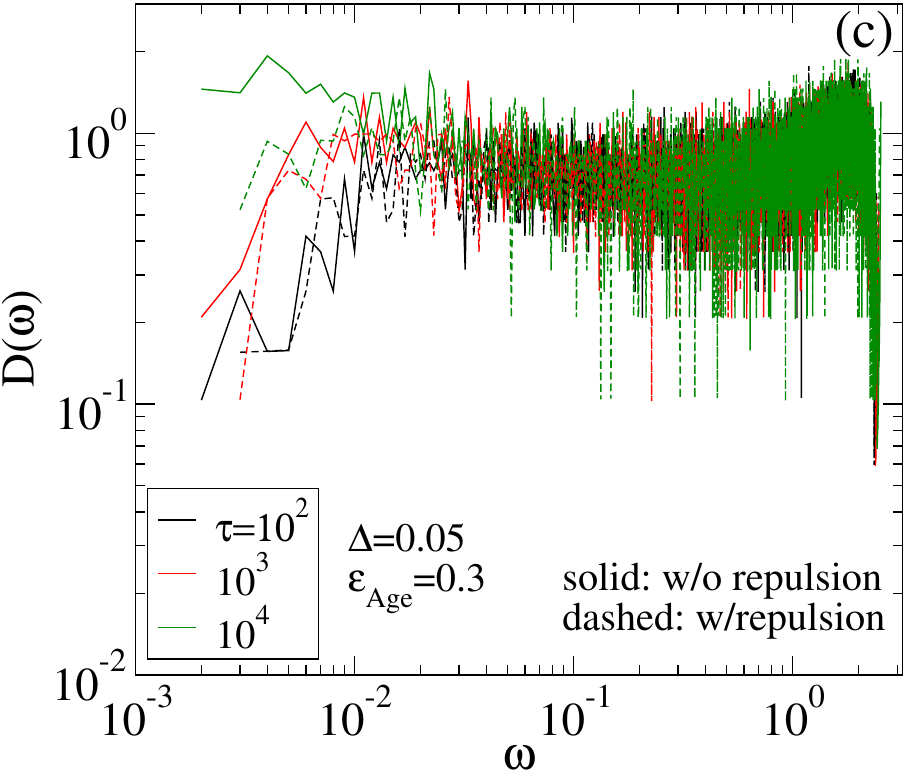}
    }
\caption{ \label{SI_fig_RegulaCoupLear} {\bf Coupled learning with angular regularization} The energy (a) and training error (b) as a function of the number of training cycles with and without repulsion for different training amplitude. (c) The density of states at different stages of training with and without repulsion. Here, $N=200$, $\Delta Z=0.03$ $\Delta=0.05$.}
\end{figure*}

Fig. \ref{SI_fig_RegulaCoupLear}  shows the results for coupled learning with angular regularization (with $\theta_c=0.15$ and $k_\theta=0.1$), discussed in the main text. Also here, the energy decays approximately as a  power-law with the number of training cycles, $U\sim \tau^{-0.5}$ with and without repulsion (see  Fig. \ref{SI_fig_RegulaCoupLear}(a)). In Fig. \ref{SI_fig_RegulaCoupLear}(b) we show the variation of error with the number of training cycles for different strain amplitudes. In the presence of repulsion, the rate of error convergence is accelerated compared to cases without repulsion. Moreover, for a large strain amplitude ($\epsilon_{Age}=0.4$) training without repulsion does not converge, while with repulsion it does converge. This demonstrates the effectiveness of angular regularization. 

Lastly, in Fig. \ref{SI_fig_RegulaCoupLear}(c) we present the density of states with repulsion and without. In the case of training with repulsion, there is a noticeable suppression of spurious low-frequency modes, while without repulsion, an increasing number of spurious modes emerge as the training progresses. Such observation is consistent with what we observed for the training rule based on plasticity. Hence, the proposed angular regularization applies to a broader set of training rules.

\end{document}